\documentclass[twocolumn,tighten,times,astrosymb,floatfix]{aastex631}

\usepackage{
	amsmath,
	amssymb,
	amsthm,
	booktabs,
        breqn,
	epsfig,
	graphicx,
	grffile,
	import,
	morefloats,
	microtype,
	natbib,
	subfigure,
	ulem,
	url,
	verbatim,
	wasysym,
	xcolor,
	xspace,
}
\usepackage{savesym}
\savesymbol{tablenum}
\usepackage{siunitx}
\restoresymbol{SIX}{tablenum}
\usepackage[version=4]{mhchem}
\usepackage[mathlines]{lineno}

\DeclareSIUnit{\jy}{Jy}
\DeclareSIUnit{\beam}{beam}
\DeclareSIUnit{\kms}{\kilo\meter\per\second}

\shorttitle{ALMA HD 166191}
\shortauthors{Worthen et al.}

\begin{document}

\title{\Large
Probing the era of giant collisions: millimeter observations of the HD 166191 system
}

\correspondingauthor{Kadin Worthen}
\email{kworthe1@jhu.edu}

\author[0000-0002-5885-5779]{Kadin Worthen}
\affiliation{William H. Miller III Department of Physics and Astronomy, John's Hopkins University, 3400 N. Charles Street, Baltimore, MD 21218, USA}

\author[0000-0002-8382-0447]{Christine H. Chen}
\affiliation{Space Telescope Science Institute, 3700 San Martin Drive, Baltimore, MD 21218, USA }
\affiliation{William H. Miller III Department of Physics and Astronomy, John's Hopkins University, 3400 N. Charles Street, Baltimore, MD 21218, USA}

\author[0000-0002-4803-6200]{A. Meredith Hughes}
\affiliation{Department of Astronomy, Van Vleck Observatory, Wesleyan University, 96 Foss Hill Drive, Middletown, CT 06459, USA}

\author[0000-0002-4267-093X]{Brandon C. Johnson}
\affiliation{Department of Earth, Atmospheric, and Planetary Sciences, Purdue University, 550 Stadium Mall Drive, West Lafayette, IN 47907, USA } 
\affiliation{Department of Physics and Astronomy, Purdue University, 525 Northwestern Avenue, West Lafayette, IN 47907, US}

\author[0000-0002-4388-6417]{Isabel Rebollido}
\affiliation{European Space Agency (ESA), European Space Astronomy Centre (ESAC), Camino Bajo del Castillo s/n, 28692 Villanueva de la Cañada, Madrid, Spain
}

\author{Diego E. Garcia}
\affiliation{Department of Physics, Middlebury College, 14 Old Chapel Road, Middlebury, VT 05753 USA}

\author{Jamar Kittling} 
\affiliation{Kavli Institute for Particle Astrophysics and Cosmology, Stanford University, Stanford, CA 94305, USA}

\author[0000-0002-9548-1526]{Carey M. Lisse}
\affiliation{Johns Hopkins University Applied Physics Laboratory, 11100 Johns Hopkins Rd, Laurel, MD 20723, USA}

%\author{Cicero X. Lu}
%\affiliation{Gemini Observatory/NSF’s NOIRLab, 670N. A’ohoku Place, Hilo, HI 96720, USA}

\begin{abstract}
We present non-simultaneous ALMA band 7 and SMA observations of the HD 166191 disk, which was recently thought to have a collision in its terrestrial planet zone. Both observations detect dust continuum emission and the ALMA observations detect the $^{12}$CO J=3-2 line from the circumstellar disk. We do not detect SiO, a potential indicator of giant collisions, but place a limit on the total SiO mass in the system. Unlike previously observed in the infrared, we do not find evidence for variability at millimeter wavelengths when comparing the ALMA continuum observations from 2024 to the pre-collision SMA observations from 2014. We perform modeling of the CO and continuum visibilities and find that both the CO and dust are marginally spatially resolved and are contained to within 20 au from the central star. The modeling of the CO suggests that the outer regions of the disk are gas rich, although further observations are needed to confirm the total gas mass. The evolutionary state of this system has been debated in the literature, and our observations, while not definitive, are generally consistent with the idea that this disk is similar to an evolved protoplanetary or transition/hybrid disk. This could suggest that collisions in the terrestrial planet zone of HD 166191 are occurring while the disk is in a transitional phase, where the inner few au are depleted of gas. This makes HD 166191 an important object for understanding the transition between protoplanetary and debris disks and the stages at which collisions occur.

\end{abstract}

\keywords{
    planet formation ---
    debris disks ---
    circumstellar matter 
}

%%%%%%%%%%%%%%%%%%%%%%%%%%%%%%%%%%%%%%%%%%%%%%%%
%			       Begin Body Text
%%%%%%%%%%%%%%%%%%%%%%%%%%%%%%%%%%%%%%%%%%%%%%%%

\section{Introduction}\label{sec:Introduction}
The formation of planets is thought to occur within the circumstellar disks of gas and dust around young stars \citep{Mamajek09, Williams11}. Theoretical simulations predict that the final growth of terrestrial planets depends on  collisions that merge planetary embryos in a gas depleted environment after the bulk gas of the protoplanetary disk has been cleared \citep{Agnor1999, Asphaug06, Raymond14}. These collisions are predicted to release debris into the circumstellar environment where it could have observable signatures at infrared wavelengths \citep{Jackson12}. Time monitoring at infrared wavelengths, as well as the dust minerology, of some disks have found evidence for ongoing giant collisions, potentially involved in the formation of terrestrial planets \citep{Lisse09,Meng14,Su19,Su22}. Identifying and classifying these systems that have ongoing collisions presents an opportunity to understand when these collisions occur in the sequence of planetary system evolution. It is currently not well constrained observationally how early destructive collisions between minor bodies can occur in planetary systems. That is, if they can begin once the primordial gas begins to be cleared or if they primarily occur in the debris disk stage of evolution, where primordial gas has dissipated and the dust is thought to be entirely of second generation (generated through collisions) \citep{Wyatt07,Hughes18}. 

The transition between the gas-dominated protoplanetary disk stage and the dust-dominated debris disk is currently not well understood \citep{Wyatt15, Michel21}. \cite{Wyatt15} proposed that hot dust in the inner regions of some transitional and protoplanetary disks may be due to ongoing collisions potentially due to the formation of rocky planets, suggesting that destructive collisions could begin in the early phase of evolution as protoplanetary disks start to transition into debris disks. Classifying the evolutionary stages of different disks to create an evolutionary sequence, especially ones that exhibit evidence of ongoing collisions would help in understanding how and when these debris generating collisions begin in the context of circumstellar disk evolution. 

Classifying individual systems to be protoplanetary disks, debris disks or transitional between protoplanetary and debris disks is not trivial as some systems have been classified in different categories by different studies (e.g. \citealt{Furlan07,Wyatt07,Schneider13,Kennedy14}). Some distinguishing factors can be the mass and composition of gas, the dust mass, and fractional infrared excess. Typically, protoplanetary disks have dust masses that are an order of magnitude greater than debris disks \citep{Panic13}. Some protoplanetary disks can have optically thick millimeter emission, which could make the dust mass difficult to constrain for some systems \citep{Villenave20,Xin23}. Further, protoplanetary disks are thought to have larger gas masses than debris disks with most of the mass in H$_2$, while these large H$_2$ reservoirs are presumably not present in debris disks \citep{Wyatt15,Matra17}.

HD 166191 is an object of interest for understanding giant collisions as it was seen to have time variability in its infrared excess indicating a recent collision \citep{Su22}. Further, its evolutionary state is currently debated in the literature (e.g. \citealt{Schneider13, Kennedy14, Su22}). HD 166191 is a late F- to early G-type star at a distance of 101 pc \citep{Schneider13, Kennedy14, Gaia2020}. Its disk has been classified both as a two component extreme debris disk and a transition disk that is thought to be gas-rich. \cite{Kennedy14} suggested that HD 166191 may be in the process of transitioning from a protoplanetary disk into a debris disk based on their SED model of its large fractional infrared excess ($\sim0.1$), lack of accretion indicators, and the large amounts of hot dust in the system that exhibits variability. \citep{Schneider13, Su22} however suggest that the system is an extreme debris disk with two separate dust populations based on the ability for a 2 component model to explain the system as well as the previous lack of gas detections. 

The age of HD 166191 has also been uncertain in the literature, with \cite{Kennedy14} suggesting that it is co-moving with the Herbig star HD 163296 and at an age of $\sim$ 5 Myr. HD 166191 was later found to be a member of a small moving group on the periphery of the Scorpius-Centaurus OB association \citep{Potranov18}, but the group was thought to have a median age of 10 Myr. More recent analyses characterized this group of stars including HD 166191 and HD 163296 and find an isochronal age of 4.2$^{+1.2}_{-1.5}$ Myr \citep{Hunt23,Hunt24}, further suggesting an age for HD 166191 of less than 10 Myr. Protoplanetary disk lifetimes are observationally thought to be between $\sim$2-8 Myr \citep{Mamajek09,Williams11,Michel21}, placing HD 166191 in an age range that could correspond to the later stages of protoplanetary disk evolution. 

 The age and evolutionary state of HD 166191 is of interest as \cite{Su22} discovered that the infrared excess of HD 166191 at 3.5 and 4.6 $\mu$m observed with \textit{Spitzer} was time variable, with an increase in flux by a factor of 2 from 2018-2020. Those observations did not show if or when the infrared excess decreased to its steady state value as the \textit{Spitzer} mission ended in 2020. At a similar time as the infrared brightening, \cite{Su22} also found that there were two occultation events that blocked $\sim 80\%$ of the stellar light at optical wavelengths. These phenomena were thought to be due to a collision between rocky bodies in the terrestrial planet zone of the system that released a clump of dust which transited in front of the star. The temperature of the dust and the period between the two occultation events suggest that this potential collision occurred at a stellocentric distance of 0.6 au, however, only two transits of the dust clump were observed with a period between the transits of 142 days. The first occultation occurred between 2017 and 2018. This recent discovery of a potential collision between rocky bodies in the terrestrial planet zone of HD 166191 \citep{Su22}, as well as its ambiguous evolutionary state, makes this a prime target to study how early these types of collisions occur during the sequence of planetary system evolution.

Here we present SMA and ALMA band 7 observations of the HD 166191 system in Section 2. With these observations, we detect dust continuum emission as well as CO emission while putting an upper limit on the amount of SiO in the system. We perform radiative transfer modeling on the visibility data of the dust and CO to determine the properties of the disk in Section 4. In Section 5 we discuss what our results imply about the HD 166191 system and also place the proposed collision in the system in context with the properties of the disk determined from our ALMA observations.

\section{Observations and Data Reduction}\label{sec:Observations}

\subsection{SMA Observations}
\begin{figure}
    \centering
    \includegraphics[scale=0.55]{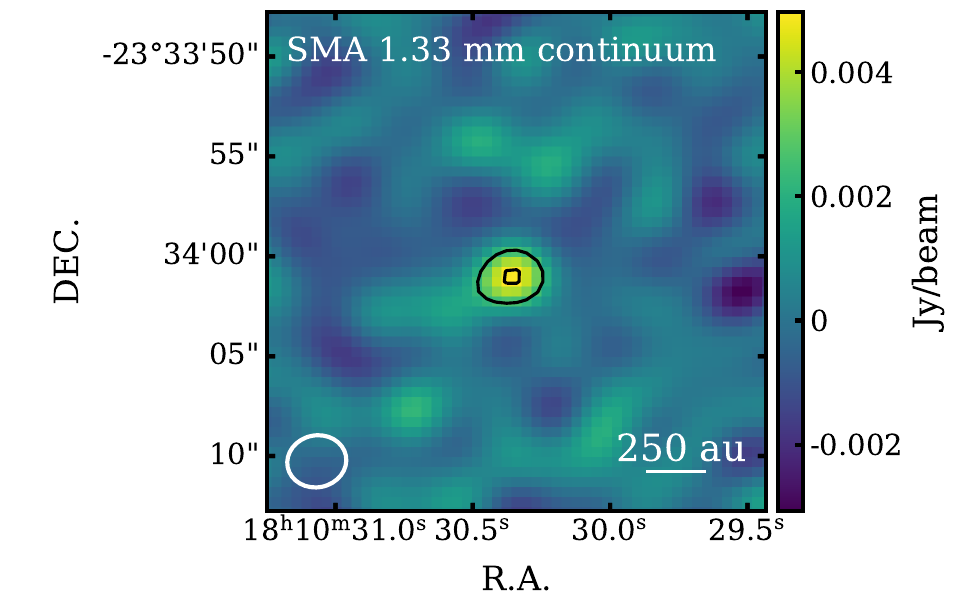}
    \caption{1.33 mm continuum SMA image of HD 166191. Contours are drawn in levels of 3 and 6 $\sigma$. The image is oriented such that north is up and east is left.}
    \label{fig:SMA_cont}
\end{figure}

HD 166191 was observed with the Submillimeter Array (SMA) \citep{Ho04} on May 14, 2014 (project code 2014A-S053, P.I. A. M. Hughes) with a total on source exposure time of 4.29 hours. Seven elements of the array were used in the compact configuration with 21 baselines. The SMA visibilities were reduced and calibrated with the MIR software package \footnote{\url{http://www.cfa.harvard.edu/~cqi/mircook.html}}. Titan and Neptune were used as flux calibrators for HD 166191 while the quasars 3c279 and 3c454.3 were used as bandpass calibrators. Quasars 1733-130 and 1832-206 were used as gain calibrators. Observations of the continuum were taken for each sideband with 4 GHz bandwidth centered around the local oscillator frequency of 225.5 GHz. We converted the calibrated SMA visibilities into measurement sets using pyuvdata \citep{Pyuv17,Pyuv25} to image them with the CASA software \citep{CASA22}. These SMA visibilities were imaged using CASA task tclean with Briggs weighting and a robust parameter of 0.5 resulting in an image with a final synthesized beam of 2.97$\times$2.60''. The final continuum image at 1.33 mm is shown in Figure \ref{fig:SMA_cont}, giving a detection of the spatially unresolved dust continuum at a $6\sigma$ level at 1.33 mm. The flux density from the SMA continuum image at 1.33 mm is measured to be 5.3$\pm$1.0 mJy/beam. 

We also use CASA to create a spectral cube from the SMA observations that covers the CO J=2-1 line at 230.538 GHz. We used a robust parameter of 1.0 to create the CO J=2-1 spectral cube from the SMA observations which had a spectral resolution of 1.0 km/s. We do not detect CO J=2-1 emission in the SMA data and place a 3$\sigma$ upper limit on the integrated line flux of 1.0 Jy km/s by integrating the SMA CO data cube from velocities of -16 km/s to 25 km/s. The channel maps of the SMA spectral cube covering those velocities is shown in the appendix.

\subsection{ALMA Observations}

\begin{figure}[!htb]
    \centering
    \includegraphics[scale=0.5]{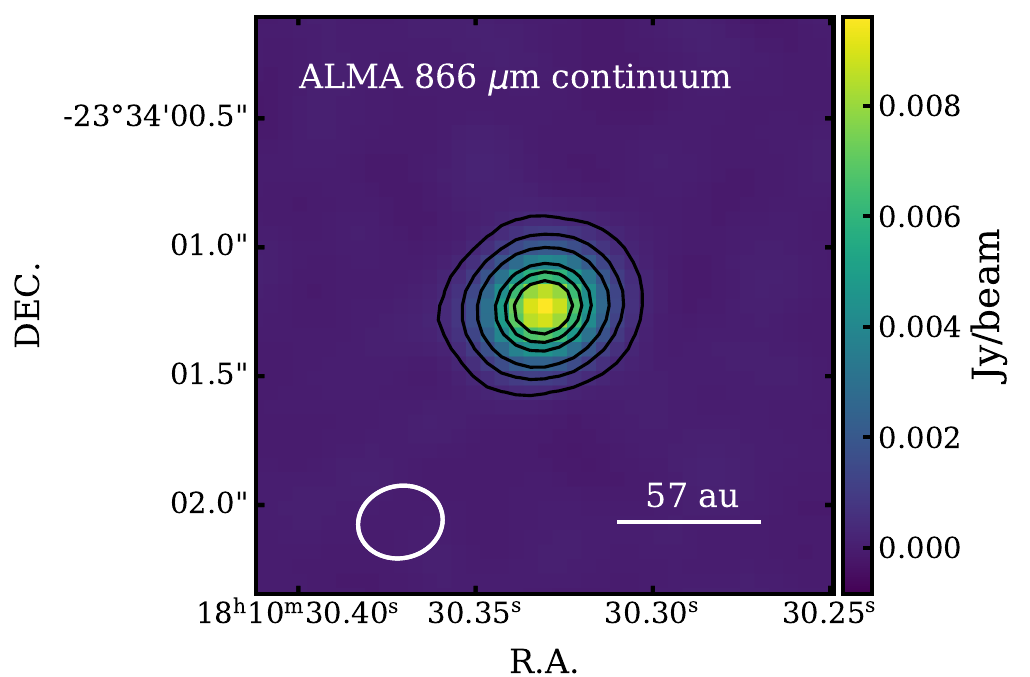}
    \caption{Continuum image of HD 166191 at 866 $\mu$m. The white ellipse in the bottom left corner shows the synthesized beam. The contours shown are at levels of 5, 15, 30, 60, 80, and 100$\sigma$ where $\sigma=$ 70 $\mu$Jy/beam. }
    \label{fig:Cont_image}
\end{figure}

HD 166191 was observed with the Atacama Large Millimeter/submillimeter Array (ALMA) in Band 7 (project 2023.1.00691.S, PI: K. Worthen) on May 20th and 24th, 2024 with a total integration time across both nights of 261 minutes. The precipitable water vapor during the observations was $\sim$0.515-0.600 mm. Quasar J1924-2914 was used as a bandpass and flux calibrator while J1820-2528 was used as a phase calibrator and the data were calibrated by ALMA staff. The correlator setup included four spectral windows, one for the SiO J=8-7 line centered at 347.35418 GHz, one for the CO J=3-2 at 345.79599 GHz, and two for continuum emission at 335.50 GHz and 334.10 GHz. The SiO and CO spectral windows both had a spectral resolution of 1.953 MHz (1.7 km/s) and bandwidths of 937.50 MHz ($\sim$ 810 km/s). The two continuum channels had bandwidths of 1875.0 MHz.

Using these observations, we produce a dust continuum image as well as spectral cubes to search for gas emission using the CASA software \citep{CASA22}. We image the continuum using the CASA task tclean. We use the line free channels of the CO and SiO spectral windows combined with the continuum dedicated windows to produce a dust continuum image at 866 $\mu$m. We produce the continuum image using Briggs weighting with a robust parameter of 0.5, yielding a synthesized beam size of 0.33''x 0.28''. The final continuum image has a pixel size of 0.056''. The continuum image at 866 $\mu$m of the HD 166191 system is shown in Figure \ref{fig:Cont_image}. To image the molecular line emission, we first remove the continuum in the visibility domain by fitting the continuum in the line-free channels and subtracting this continuum fit from all spectral channels using the CASA task UVCONTSUB. We use the same tclean parameters (Briggs weighting with robust=0.5) to produce spectral cubes for the SiO and CO lines. 
 \begin{figure*}[!htb]
    \centering
    \includegraphics[scale=0.85]{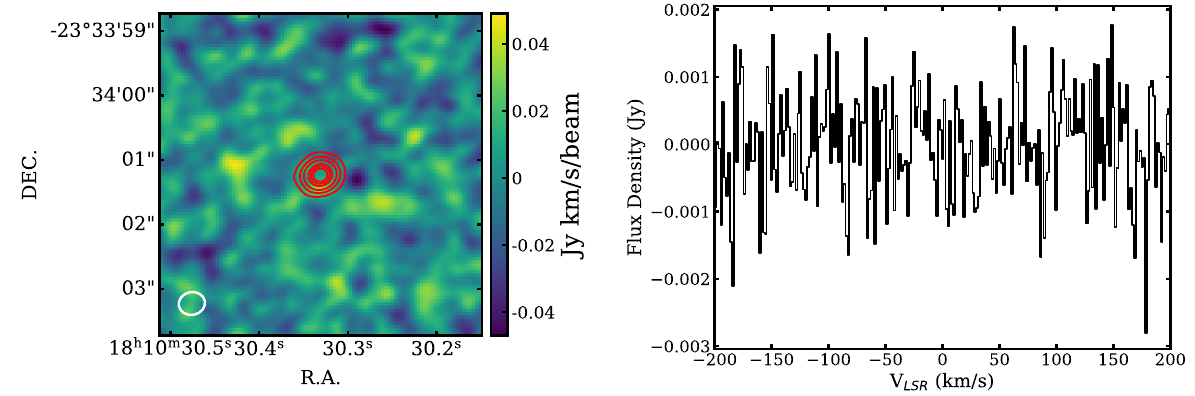}
    \caption{Left: moment 0 map of the SiO cube integrated from -50 to 50 km/s. The red contours show the dust continuum  and are at the same levels as in Figure \ref{fig:Cont_image}. There is no clear detection of SiO at the position of the continuum detection. Right: extracted spectrum from the SiO cube showing a non-detection of SiO emission towards HD 166191.}
    \label{fig:SiO_non_detection}
\end{figure*}

We search for SiO emission, a signature of giant collisions \citep{Lisse09,Johnson16}, in our spectral cubes but find no significant emission in the channels of the SiO spectral cube. We generate another SiO spectral cube using a robust parameter of 2.0, to increase the sensitivity to SiO emission at the expense of angular resolution, and still find no detection of SiO in the channels. We extract a spectrum from the SiO cube by using the dust continuum as a spatial filter and summing only pixels that are within the 5$\sigma$ contour of the dust continuum image, as shown in Figure \ref{fig:Cont_image}. We also integrate the SiO cube from velocities of -50 to 50 km/s using the SpectralCube package \citep{SpectralCube} to produce a moment 0 map. We chose this velocity range because the Keplerian velocity around HD 166191 (1.6 M$_{\odot}$) at 0.6 au, where the collision is thought to have occurred \citep{Su22}, is 48 km/s. This range thus captures the entire range of velocities expected for SiO in orbit at that distance from HD 166191. The spectrum from the SiO cube along with the moment 0 map, which both show non-detections of SiO, are displayed in Figure \ref{fig:SiO_non_detection}.

\begin{figure*}[!htb]
    \centering
    \includegraphics[scale=0.55]{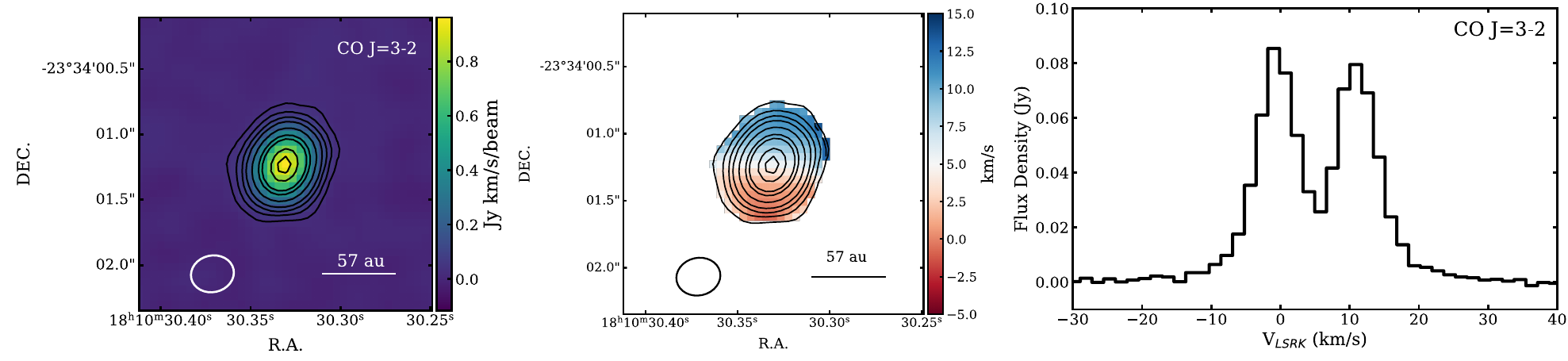}
    \caption{Left: moment 0 map of the CO J=3-2 line from the HD 166191 system. Contours are at levels of 3, 6, 10, 15, 25, 40, 50, 75, and 90 $\sigma$ of the moment 0 map where $\sigma=0.01$ Jy km/s/beam. The white oval represents the synthesized beam. Middle: moment 1 map showing the CO velocity around HD 166191. The contours are the same as the left figure. Right: Extracted CO spectra of HD 166191 showing a double peaked structure indicative of gas in Keplerian velocity around the star.}
    \label{fig:CO_gal}
\end{figure*}

In the CO spectral cubes (robust=0.5), we do find that there is emission in the channel maps. We similarly used the SpectralCube package to produce moment 0 and moment 1 maps of the CO line. The CO data cube was integrated over frequencies where there was a 3$\sigma$ detection of CO emission from the disk. The moment 0 and moment 1 maps showing the detection of CO emission are displayed in Figure \ref{fig:CO_gal}. To extract a spectrum of the CO line from the CO data cube, we sum the flux from all pixels within the 5$\sigma$ contour of the CO moment 0 map. The CO spectrum is also shown in Figure \ref{fig:CO_gal}. The CO spectrum has an asymmetry between the two peaks (at a 5$\sigma$ level where $\sigma$ is the image rms per channel). There is diffuse emission seen in the channel maps at some of the same velocities as the CO in the disk. The disk is fainter at the velocities with the contamination from the diffuse emission. It is possible that the surrounding cloud is absorbing some of the CO emission from the disk creating this asymmetry, which has been seen in other millimeter observations 
 of CO in disks that are surrounded by molecular clouds (e.g. \citealt{Schaefer09,Sheenan19}). This diffuse CO emission around HD 166191 is shown in a moment 0 map, which was integrated over the velocity channels where the diffuse emission is detected, in Figure \ref{fig:diffuse_CO}. It can also be seen in the SMA channel maps in the Appendix.

\section{Results}

\subsection{Dust}
 We use the fitting tool in the CASA viewer to obtain basic parameters of the dust continuum emission from a 2D Gaussian fit. This gives an integrated continuum flux density of 11.0$\pm$0.2 mJy at 866 $\mu$m. The continuum FWHM convolved with the beam is 0.343$\pm$0.003''. This is slightly larger than beam size of 0.33'', suggesting that the continuum may be marginally resolved. This is confirmed below with the visibility modeling of the dust continuum to further explore its spatial extent. The position angle of the continuum from the 2D Gaussian fit is 164$\pm$27 degrees.  

To search for variability at millimeter wavelengths and to constrain the temperature of the dust to estimate the total dust mass in the optically thin limit, we perform blackbody fitting to the SED of the HD 166191 system. For simplicity, we use a two-component blackbody model to fit the SED to determine if the ALMA and SMA points can be explained by the same blackbody. In the literature, the SED of HD 166191, excluding millimeter data, has been fit using both a two component dust model \citep{Schneider13} as well as a continuous protoplanetary disk model \citep{Kennedy14}. We do not favor one model to explain the SED over the other here, but rather just choose the two component model for simplicity since we are mainly interested in checking for variability between the millimeter data points from 2014 and 2024. 

\begin{figure}[!htb]
    \centering
    \includegraphics[scale=0.45]{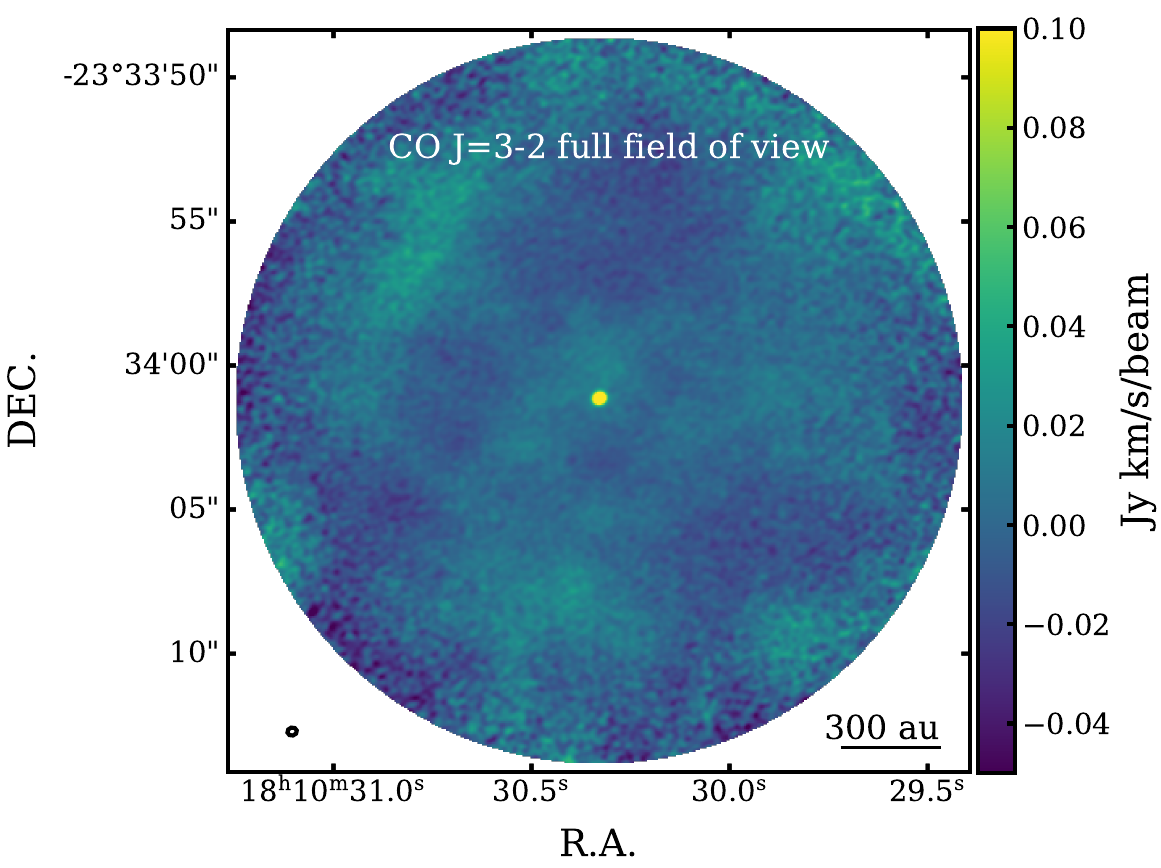}
    \caption{Moment 0 map of the entire field of view of the HD 166191 observations showing diffuse emission at the same velocities as the CO in the disk. The bright emission at the center of the image is from the disk. This moment 0 map is integrated from velocities of 7.5 to 11 km/s.}
    \label{fig:diffuse_CO}
\end{figure}

In the SED fit, we include JHK, WISE, IRAC, and \textit{Spitzer} MIPS photometry presented in \cite{Schneider13} and references therein. We also include our ALMA and SMA continuum fluxes as well as the IRTF SpeX and BASS spectra presented in \cite{Kennedy14}. For the stellar photosphere we use an BT-Settl photosphere model \citep{Allard03} with $T_{\mathrm{eff}}=6100$ K \citep{Schneider13,Kennedy14, Su22} and log(g)=3.5 \citep{Kennedy14} that we pin to the H-band magnitude of the system \citep{Cutri03}. We include a warm and cold blackbody component in our fit where the cold blackbody is a modified blackbody since the opacities of dust in disks can have a power law dependence at mm wavelengths \citep{Roberge13,Yang24}. We therefore multiply the cold blackbody component in our model by a factor of $(\frac{\lambda}{\lambda_0})^{-\beta}$ for wavelengths larger than $\lambda_0$. Here $\lambda_0$ is a reference wavelength that is left as a free parameter in the fit and $\beta$ is the emissivity power-law index that is also left as a free parameter. 

The BASS spectrum from \citep{Kennedy14} has a shape indicative of a 10$\mu$m silicate feature, so we also include a component of silicate emission in our model. The main goal of including this in the model is not to precisely characterize the dust minerology given the low S/N of the BASS spectrum ($\sim$3-15 across the spectrum), but instead to reproduce the 10$\mu$m feature and then predict the resulting 20$\mu$m silicate feature, since this will affect the \textit{Spitzer} photometry measurement at 24 $\mu$m and thus the cold modified blackbody in the model. We use a silicate emission model similar to \cite{Lu22} and \cite{Sargent09} where the emission from the silicate grains is assumed to be optically thin and can be modeled as 
\begin{equation}
    F_{\nu}(\lambda)=B_{\nu}(\lambda,T)\sum_j(a_j\kappa_j(\lambda))
\end{equation}
where $F_{\nu}$ is the flux density from the silicate grains, $B_{\nu}(\lambda,T)$ is the Planck function at the temperature of the grains, $\kappa_j$ is the opacity (cross-section per unit mass) of dust species $j$ at wavelength $\lambda$, and $a_j$ is the mass weight of dust species $j$. The mass weight is defined as $a_j=m_j/d^2$, where $m_j$ is the mass of dust species $j$ and $d$ is the distance to the source (101 pc). We assume the silicate grains have the same temperature as the hot dust component. We initially include three dust species in our model: amorphous olivine, amorphous pyroxene, and crystalline forsterite, although we find that the 10 $\mu$m feature can be explained with contributions from just pyroxene and forsterite. $\kappa_{\nu}$ is computed using Mie theory and optical constants from \cite{Zeidler15} for the crystalline grains and \cite{Jaeger94} and \cite{Dorschner95} for the amorphous grains. For each dust species, we include 2 sizes of dust grains of 0.5 $\mu$m and 3$\mu$m in radius, similar to \cite{Sargent09}. The free parameters in our SED model are thus the temperatures of the two blackbodies, the reference wavelength $\lambda_0$, the emissivity power-law index $\beta$, and the mass fractions of the silicate species. We fit the SED model by minimizing the $\chi^2$ value between the model and data with the differential evolution algorithm to explore the parameter space \citep{Storn97}.

\begin{figure}
    \centering
    \includegraphics[scale=0.38]{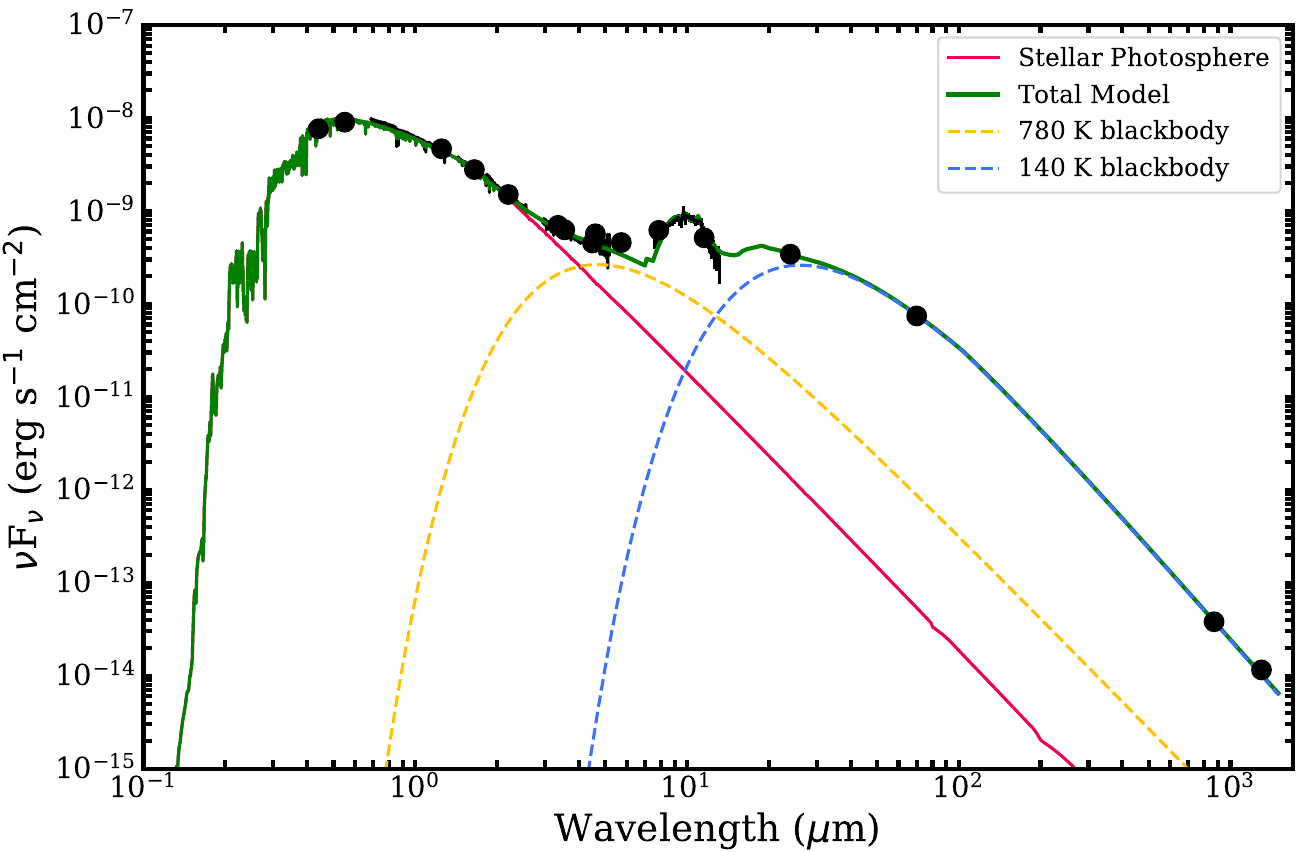}
    \caption{SED of the HD 166191 system with our best fit SED model (green). The stellar photosphere is shown in red and the hot and cold blackbody components are shown in yellow and blue respectively. The ALMA and SMA points are at 866 $\mu$m and 1300 $\mu$m respectively. Note that this SED is mainly used to search for variability at millimeter wavelengths and not as a representation of the system architecture.}
    \label{fig:SED_fig}
\end{figure}

The best fit SED model as well as the photometry and spectra of HD 166191 is shown in Figure \ref{fig:SED_fig}. The best fit temperatures of the warm and cold components are 780 K and 140 K respectively, which are similar to those found in the two component model of \cite{Schneider13}. We find a best fit references wavelength of $\lambda_0=103$ $\mu$m and an emissivity power-law index of $\beta=0.36$. The best fit masses of the 3 $\mu$m radius pyroxene grains and 0.5 $\mu$m forsterite grains are 1.3$\times10^{22}$ and 3$\times10^{20}$ g. The other sizes and species of grains included in the model fit have best fit masses that are the minimum allowed mass we allowed in the range for the fit so these grains have masses $<10^{20}$ g. This simple silicate model is able to reproduce the 10 $\mu$m feature given the S/N of the BASS spectrum. Higher S/N data in the mid-IR and a more complex silicate model would be required to precisely determine the minerology of the grains in this disk, but this is beyond the scope of the paper. We do not find evidence of variability at millimeter wavelengths, as the SMA measurement taken in 2014 and the ALMA measurement taken in 2024 are well explained by one blackbody (i.e. the 2024 ALMA measurement is not significantly different from 
a model that is consistent with the level of millimeter flux measured in 2014 before the collisional event seen with \textit{Spitzer}). 

Using the two millimeter measurements from ALMA and the SMA, we can directly compute the spectral index $\alpha$ using $\alpha$=ln($F_{\nu_1}$/$F_{\nu_2}$)/ln($\nu_1$/$\nu_2$). Using the ALMA flux density measurement of 11.0$\pm$0.2 mJy and the SMA flux density of 5$\pm$1 mJy, we find $\alpha=1.9\pm$0.6. Low values for the millimeter dust spectral index ($\alpha\sim2$) can be due to the dust emission being optically thick or the dust sizes being large (millimeter to centimeter in size) \citep{Huang18,Villenave20}. If the dust emission is optically thick, then the simple two component model of optically thin dust used to fit the SED above may not be a good representation of the system, which is what is suggested by \cite{Kennedy14}. It is, however, still useful here in showing that the millimeter dust does not show evidence of variability.

Using the temperature of the cold component of the dust (140 K) from our SED model, we can estimate the total millimeter dust mass of the disk assuming that the dust emission is optically thin. The dust mass of the disk can be estimated with the optically thin assumption using the equation
\begin{equation}
    M_d=\frac{F_{\nu}d^2}{B_{\nu}(T)\kappa_{\nu}}
\end{equation} \label{eq:Mdust}
where $M_d$ is the dust mass, $F_{\nu}$ is the millimeter flux, $B_{\nu}(T)$ is the Planck function at the temperature of 140 K derived from the SED fit, and $\kappa_{\nu}$ is the dust opacity. For the dust opacity, we use an opacity that is commonly used in the literature (e.g. \citealt{Beckwith90,Ansdell16,Michel21}) of $\kappa_{\nu}=10$ cm$^2$/g at 1000 GHz with a power-law index of $\beta=1$. At 866 $\mu$m, this gives a dust opacity of 3.44 cm$^2$/g. Using equation 2 and the measured flux of the continuum image of HD 166191 of 11.0$\pm$0.2 mJy, we calculate a dust mass of 0.107$\pm$0.003 M$_{\oplus}$. This dust mass serves as a lower limit, as radiative transfer models described in later sections, as well as the low millimeter spectral index value $\alpha$, suggest that the millimeter emission seen with ALMA can also be explained with a marginally optically thick disk model with higher dust mass.

\subsection{SiO non-detection and CO Emission}

Although these observations do not detect SiO, an indicator of hyper velocity collisions between large rocky bodies \citep{Lisse09,Johnson16}, we can place an upper limit on the amount of SiO present at the time we observed this system. We calculate the noise level (standard deviation of pixels in the 3 sigma contour of the continuum image) of the SiO cube moment 0 map to be $\sigma$=7$\times$10$^{-22}$ W/m$^2$, after correcting for correlated noise by multiplying by the square root of the number of beams in the continuum mask as in \cite{Carney19}. Assuming that SiO emission is optically thin and in LTE, we can put a limit on the SiO mass in the disk using the equation
\begin{equation}
    M_{\mathrm{SiO}}=\frac{4\pi md^2F}{fh\nu A}
\end{equation}\label{eq:gas_mass}
where $d$ is the distance to the source, $F$ is limit on the integrated line flux, $f$ is the fractional population of the upper energy state, $h$ is the Planck constant, $m$ is the mass of the SiO molecule, $\nu$ is the frequency of the transition, and $A$ is the Einstein A coefficient of the transition. Using equation 3, and a gas temperature equal to the hot dust temperature from the SED fit (780 K) we place a 5$\sigma$ upper limit of SiO mass in the HD 166191 disk of M$_{\mathrm{SiO}}<$1$\times10^{-7}$ M$_{\oplus}$.

The assumption that the SiO is in LTE may not be correct at the location where the collision is thought to have occurred (0.6 au). Using collision rates of SiO and \ce{H2} at 700 K \citep{Balanca18}, we find that the critical density of \ce{H2} as a collisional partner for the SiO J=8-7 transition is n$_{\ce{H2}}$=8.6$\times10^{6}$ cm$^{-3}$. Based on our modeling to the CO data in subsequent sections, the gas densities in the disk would be larger than that critical density in some regions of the disk at distances greater than 2 au if \ce{H2}/CO=1$\times10^{4}$. However, as indicated by our modeling, the region of the disk at 0.6 au may be depleted in gas. This would make the gas densities below the critical density in this region so the assumption of LTE may not hold. As shown in \cite{Matra15, Matra18}, non-local thermodynamic equilibrium (NLTE) conditions can lead to the fractional populations of higher rotational states like the J=8 state being lower than in LTE. If the SiO is not in LTE, then this would increase our upper limit on the total SiO mass. The exact amount that the upper limit would increase would depend on the kinetic temperature and  number density of the gas in the inner regions of the disk, which is currently unknown. \cite{Matra15}, however, shows that for CO in the Fomalhaut system, the gas mass calculated in NLTE can increase by $\sim$3 orders of magnitude compared to that in LTE.

We find that CO is detected towards HD 166191 at high significance ($\sim100 \sigma$ in the integrated moment 0 map) and appears to be spatially resolved (see Figure \ref{fig:CO_gal}). The resolved kinematics in the moment 1 map and the double peaked CO spectrum is typical for gas in Keplerian orbit around a star (e.g. \citealt{Hales22}) confirming that this CO emission arises from a disk around HD 166191. Integrating the CO emission over velocity gives a line flux of 1.3 Jy km/s. We use the fitting function in the CASA viewer to derive basic properties of the CO distribution by fitting a 2D Gaussian to the moment 0 map. In subsequent sections, we perform a more intensive modeling procedure to constrain the disk parameters by fitting radiative transfer models to the CO visibilities to compare to this simple Gaussian fit. From the Gaussian fit, we find that the CO emission in image space has a major axis FWHM convolved with the beam of 0.42$\pm$0.01'' and deconvolved from the beam of 0.30$\pm$0.02''. Given that the synthesized beam has a major axis FWHM of 0.33'', this suggests along with the kinematic profile that the CO emission is spatially resolved in our data. From this Gaussian fit, we also find that the position angle of the CO emission deconvolved from the beam is 167$\pm$3 degrees. With the Gaussian fit, we find that the minor axis (FWHM) of the CO disk convolved with the beam is 0.34$\pm$0.01'', and 0.11$\pm$0.04'' deconvolved with the beam. The FWHM of the minor axis of the synthesized beam is  0.28'', suggesting that the CO disk has a marginally resolved minor axis. Assuming that the disk is a flat circle, we can derive the inclination from the major and minor axes using cos($i$)=$a/b$, where $a$ is the minor axis and $b$ is the major axis. This gives an inclination of 69$^{+9}_{-11}$ degrees. Note that this inclination derivation neglects any vertical structure of the disk. Forward modeling of ALMA continuum images has shown there to be a degeneracy between the disk vertical structure and its inclination (e.g. \citealt{Hales22,Worthen24}).

\section{Radiative Transfer Modeling}

\subsection{Dust Modeling}
We perform radiative transfer modeling to the continuum visibilities to constrain the properties of the HD 166191 dust disk. The continuum visibilities were obtained by first selecting channels that do not include CO emission and then averaging the visibilities in time by 60 seconds. We then further average these spectral channels and windows into one set of continuum visibilities that we use to compare with a radiative transfer model. For the dust modeling, we use DISC2RADMC \citep{Marino22}, and we assume a Gaussian surface density for the dust with the central radius, the standard deviation of the Gaussian (a proxy for the disk width), and the total dust mass as free parameters. DISC2RADMC uses RADMC-3D \citep{Dullemond12} to compute the radiative transfer and produce synthetic images given an input of the disk surface density. We use the same dust composition and size distribution, a power-law size distribution with a power-law index of -3.5 and a minimum grain size of 1 micron and a maximum grain size of 1 cm, as in \cite{Marino18, Hales22}. The dust temperature as a function of distance from the star is computed with RADMC-3D using an effective temperature for the central star of 6100 K. We assume the disk has a Gaussian vertical distribution with a constant aspect ratio of $h/r=0.1$, a value similar to those found in other disks \citep{Matra25}. At the current resolution of the data, the vertical structure of the disk is not well resolved and a value has to be assumed. We do find that changing the scale height of the dust has an effect on the resulting flux and thus the dust mass determined from the model, so given the lack of a constraint on the scale height of the disk, the absolute value of the dust mass determined by the models should be viewed with some skepticism as an incorrect assumption for the vertical structure of the disk could lead to an incorrect dust mass. Given the lack of resolved structure in the continuum image, we fix the position angle and inclination to those which were determined from the modeling of the CO visibilities (see below). We take the continuum models generated from DISC2RADMC and transform the image into visibilities sampled at the same UV locations as the observations using GALARIO \citep{galario18} before comparison with the visibility data.  

\begin{figure*}[!htpb]
    \centering
    \includegraphics[scale=0.6]{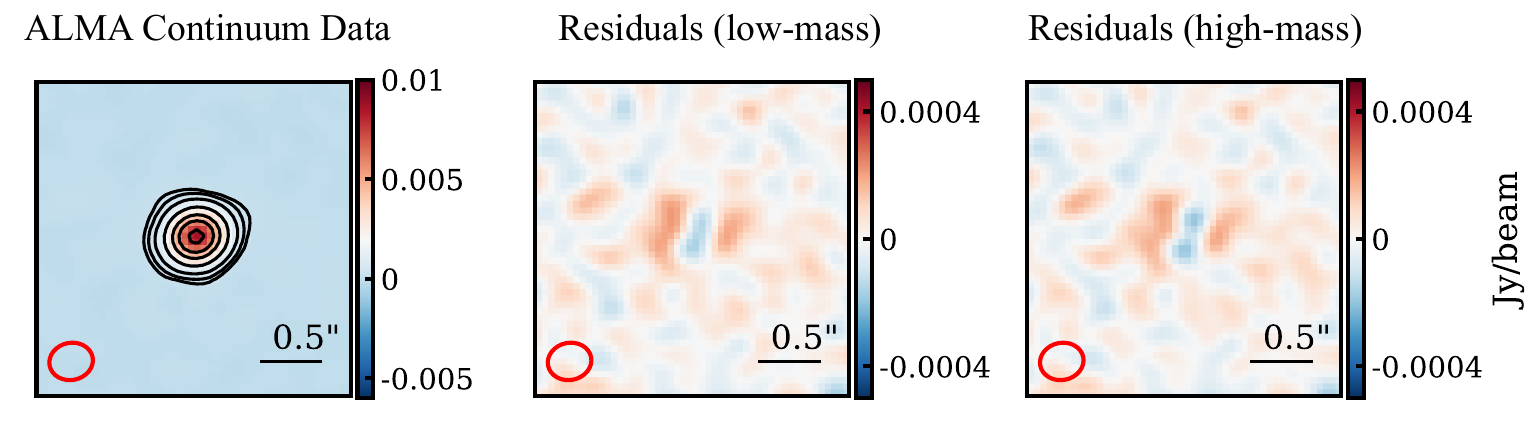}
    \caption{Continuum image and residuals after subtracting the best fit models. \textbf{Left:} dust continuum image with contours showing 3, 5, 10, 25, 50, 75, and 110$\sigma$. \textbf{Middle:} residual image after subtracting the low dust mass model visibilities. There are no contours because there are no residuals greater than 3$\sigma$. There appears to be some structure in the residuals, but it is at the 2$\sigma$ level. \textbf{Right:} same as the middle panel but for the high dust mass continuum model. }
    \label{fig:dust_resids}
\end{figure*}

Before fitting to the visibilities, we reweight the visibilities using the CASA statwt task and then scale the weights on the visibility measurements such that the reduced chi-squared value of the best fit model is 1, as in \cite{Marino21} and \cite{Hales22}. These sets of visibilities and weights are then used to compare with the continuum image from DISC2RADMC. We use the emcee package \citep{emcee13} to run a Monte Carlo Markov Chain (MCMC) to infer the posterior distributions on each free parameter.  We use a Gaussian likelihood function in the MCMC and we run the MCMC for 2000 steps with 50 walkers. We also include offsets in right ascension and declination as free parameters (dRA and dDec).

\begin{figure}
    \centering
    \includegraphics[scale=0.5]{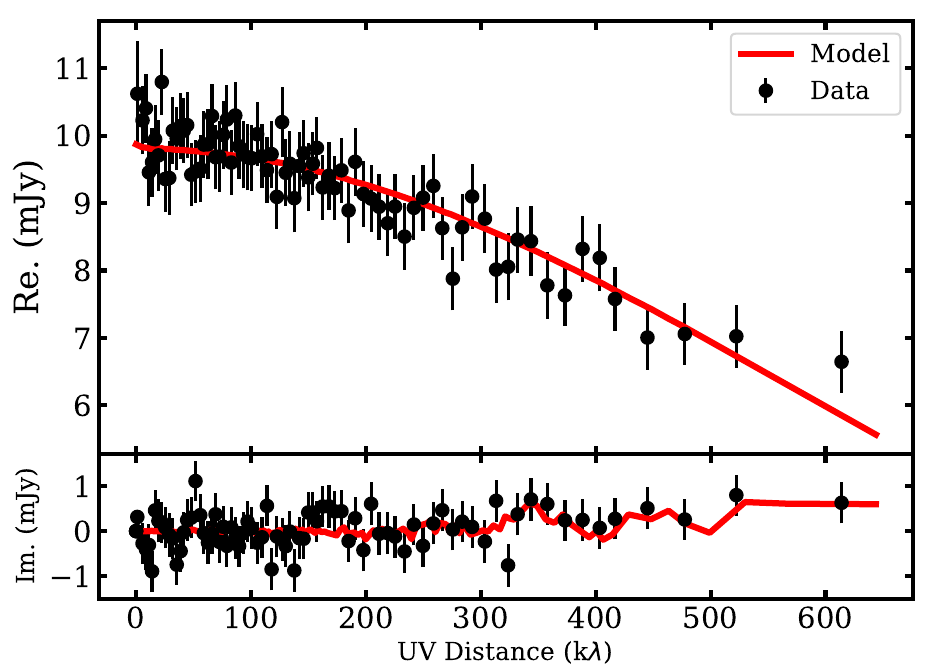}
    \caption{\textbf{Top:} real visibilities as a function of UV distance for the dust continuum of HD 166191 (black). The red line shows the real visibilities of the best fit optically thick dust model. The visibilities here are binned by a factor of 1000. \textbf{Bottom:} imaginary visibilities as a function of UV distance for the HD 166191 continuum (black). The imaginary component of the best fit continuum model shown is shown in red. }
    \label{fig:real_vis}
\end{figure}

In performing the radiative transfer modeling to the continuum visibilities, we find that we can reproduce the observations with both an optically thin and a marginally optically thick continuum disk model (peak optical depth of $\sim$2.4), and both give equally good fits to the ALMA observations (see Figure \ref{fig:dust_resids}). In each case, due to computational expense, the thermal profile of the disk was computed separately before running the model fitting, so the optically thin and optically thick models were fit independently to the visibility data. The residual images of the low dust mass and high dust mass models are shown in Figure \ref{fig:dust_resids}, demonstrating that both models can reproduce the observations to the same fidelity. The real visibility profile as a function of UV distance (after deprojecting by the disk P.A. and inclination) along with the real visibilities of the best fit moderately optically thick disk model are shown in Figure \ref{fig:real_vis}. This shows that the real visibility profile as a function of projected UV distance is not flat, indicating that the disk is marginally resolved by our ALMA observations. 

The posterior distributions for the dust model are shown in the Appendix. We find that the central radius and the disk width are correlated and there is a large uncertainty on these parameters in the low mass model, which is not surprising given the resolution of the observations and the size of the disk. This model fitting does appear to constrain the outer disk radius however to be within a stellocentric distance of 20 au. 100 randomly drawn samples of the dust and CO surface density profiles (determined from fitting to the CO data described below) from the posterior distributions are shown in Figure \ref{fig:Surface_dens}. While the observations appear to place a constraint on the dust outer radius, the dust inner radius is not well constrained at the current resolution of the observations.

 \begin{table}[!htbp]
     \centering
     \caption{Dust model parameters}
     \begin{tabular}{c c c c}
        
        \hline
        Parameter& Low Mass Model  & High Mass Model  & Prior \\
          \hline
          log($M_{\mathrm{dust}}$) (M$_{\oplus}$)& -0.04$^{+0.01}_{-0.01}$&0.568$^{+0.003}_{-0.003}$ &(-2, 2)\\
          $R_c$ (au)& 1.95$^{+3.50}_{-1.35}$&1.11$^{+1.21}_{-0.67}$ &(0.2, 30)\\
          $\sigma_R$ (au) & 5.32$^{+0.55}_{-2.21}$&5.61$^{0.30}_{-0.50}$ &(0.1, 10)\\

          \hline

     \end{tabular}
     
                \begin{minipage}{9 cm}
    \vspace{0.2cm}
    
     \textbf{Note.} The table shows the best fit values for both the low dust mass model that is optically thin and the high dust mass model that is marginally optically thick. 
    \end{minipage}
     \label{tab:Dust_params}
 \end{table}

\subsection{CO Modeling }
To constrain the parameters of the gas disk around HD 166191, we perform radiative transfer modeling that we compare to the continuum subtracted visibilities of the CO line. We also use the DISC2RADMC package \citep{Marino22} to perform the radiative transfer calculation and produce synthetic CO J=3-2 spectral cubes to compare to the observations. This package computes model spectral cubes of CO emission given an input of the disk surface density and stellar mass and gas temperature. We use a stellar effective temperature of 6100 K and a stellar mass of 1.6 M$_{\odot}$ \citep{Su22}. We assume that the CO orbits at the Keplerian velocities around HD 166191. For the surface density of the CO, we use the \cite{Lynden-Bell74} profile, which is a power-law disk with an exponential decay at larger radii. This surface density profile as a function of radius is given as
\begin{equation}
    \Sigma(r)=\Sigma_0\left(\frac{r}{R_c}\right)^{-\gamma}\mathrm{exp}\left[-\left(\frac{r}{R_c}\right)^{2-\gamma}\right]
\end{equation}
where $R_c$ is the radius where the surface density starts to decay exponentially, $\gamma$ is the power law exponent and determines how quickly the surface density decays past $R_c$, and $\Sigma_0$ is directly proportional to the total CO mass in the disk. We set the surface density of the CO to be zero below an inner radius of $R_{in}$. 

\begin{figure}
    \centering
    \includegraphics[scale=0.45]{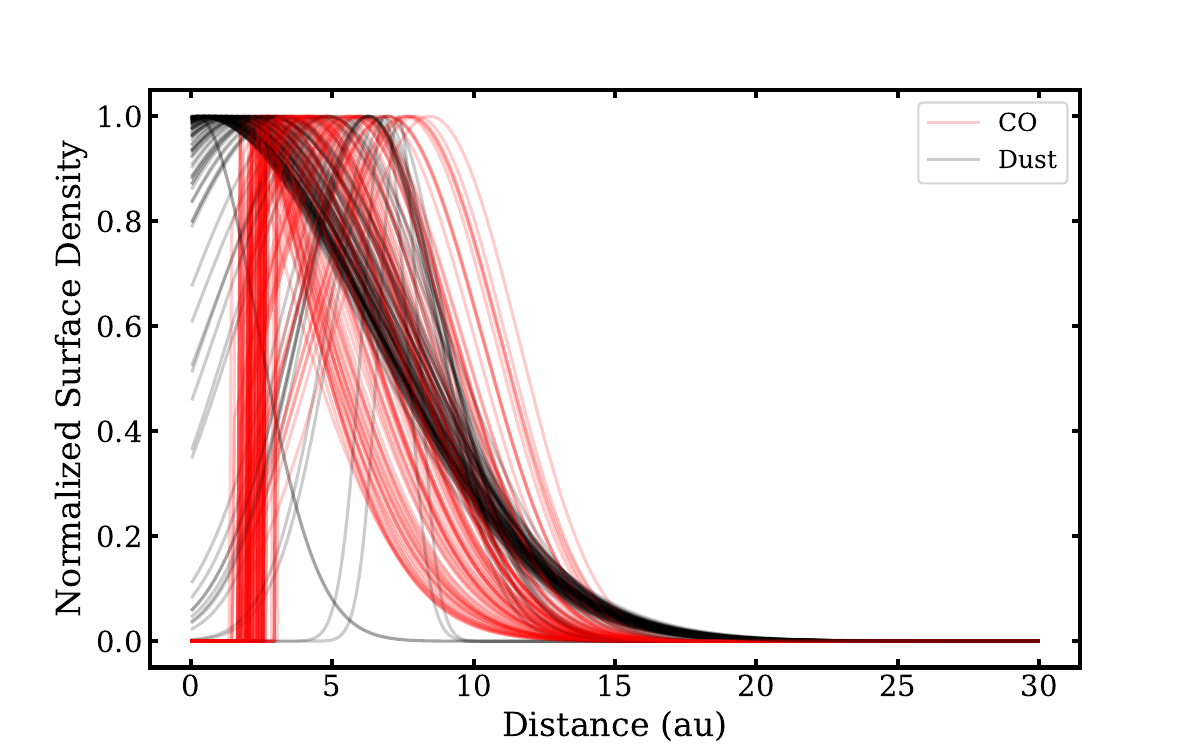}
    \caption{Normalized surface density as a function of stellocentric distance from the CO (red) and the dust (black) modeling of the HD 166191 ALMA visibilities. Each curve represents a randomly drawn sample of the posterior distributions for the CO and dust. Both the CO and the dust appear to be constrained to within 20 au. }
    \label{fig:Surface_dens}
\end{figure}

 In our model, RADMC-3D assumes that the rotational levels of CO are populated in LTE and we use the temperature profile of the dust as a proxy for the temperature of the CO. The temperature of the dust as a function of distance from the star is calculated by RADMC-3D using the same dust species as \citep{Marino18}.  We then scale this dust temperature profile, as a function of distance from the star, and use it as the excitation temperature of the CO. The scaling factor of the temperature is left as a free parameter in the model fitting. This allows the temperature of the CO to vary, but the temperature profile as a function of radius remains constant and is proportional to r$^{-1/2}$. Given that the disk's minor axis is only marginally resolved at the angular resolution of the data, we assume the gas disk has a Gaussian vertical distribution and that the scale height is given by the hydrostatic equilibrium scale height under the assumption that the mean molecular weight of the gas is $\mu=2.3$. We do find that changing the vertical structure of the disk has an effect on the CO line flux from the model, so it is possible that the best fit CO mass from our model is biased by the choice of vertical structure and thus the absolute value of the total CO mass derived from the model fitting should be viewed with some skepticism. We also test a mean molecular weight of $\mu=14$, as expected for secondary gas dominated by C and O \citep{2017MNRAS.469..521K}. This leads to a smaller scale height of the gas disk which requires a larger gas mass to produce the observed line flux. The assumption of $\mu=2.3$ therefore does not bias the results to higher CO masses. The choice of vertical structure may also affect the derived disk inclination. Previous studies have shown that disk scale height and inclination can be degenerate when modeling ALMA continuum data and models that are less inclined require smaller scale heights to reproduce the observed disk minor axis \citep{Hales22,Worthen24}. Choosing a vertical structure for the CO around HD 166191 that has a smaller scale height could lead to the derived inclination being lower. However, we also have the kinematic profile of the CO from the resolved emission line, which has been shown to possess information about the inclination of a disk (e.g. \citealt{Schneidermann21}). The best fit inclination of the CO is therefore not just derived from the disk aspect ratio with the radiative transfer modeling, but also from the velocity profile of the gas, however, higher resolution observations would ultimately be necessary to break the degeneracy between the vertical structure of the disk and the inclination.

 \begin{figure*}[!htbp]
    \centering
    \includegraphics[scale=1]{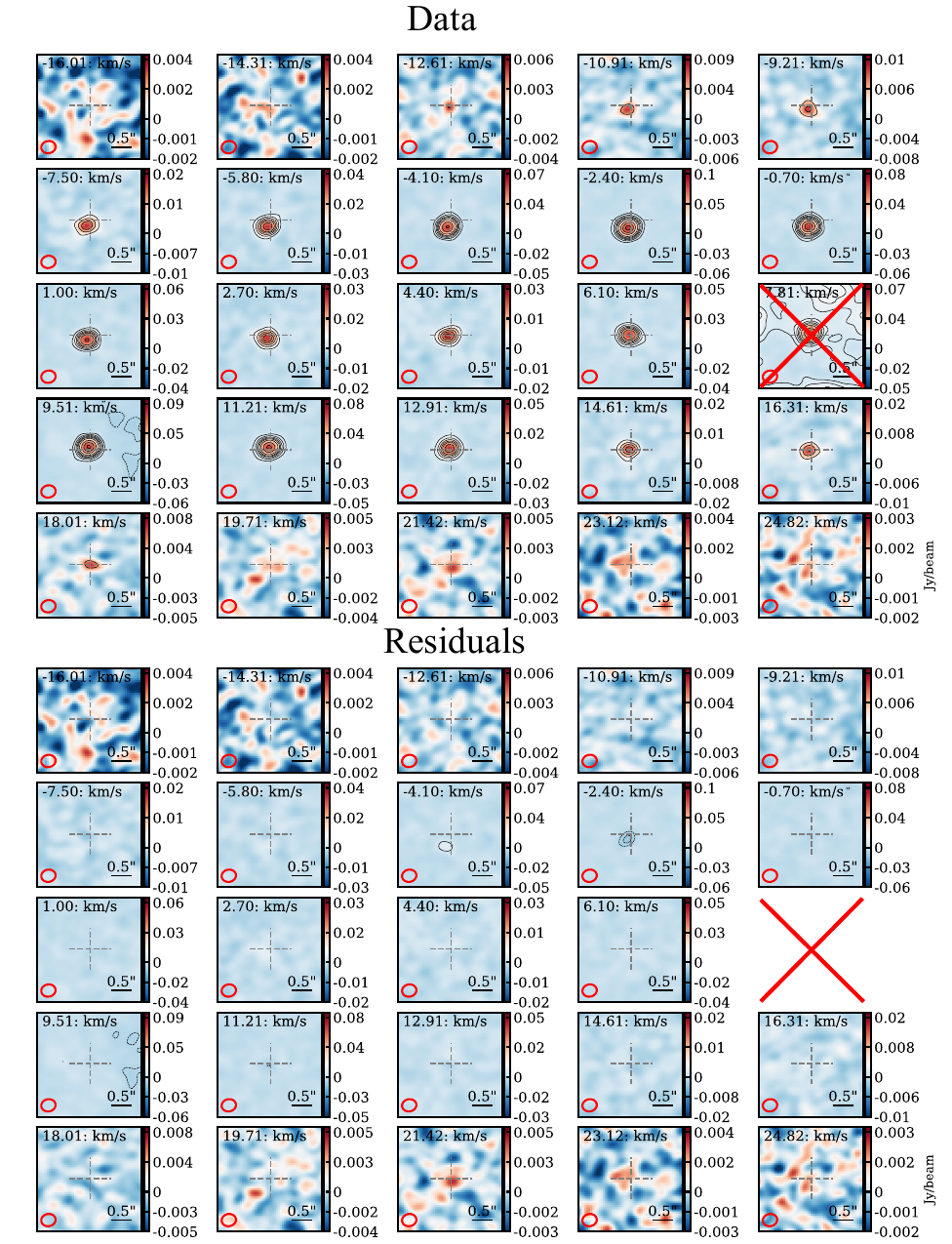}
    \caption{CO channel maps for HD 166191 are shown as the top 25 images under the label data. The gray cross denotes the central position of the CO moment 0 map. Contours are drawn in levels of 3, 6, 9, 12, 15, 20, 30, 40, 50, and 60 $\sigma$. Negative contours are shown by dashed lines and positive contours are shown by solid lines. The gray dashed cross shows the star's position. The red ellipse shows the synthesized beam. The red `x' shows the channel with contamination that was not included in the fit. The bottom 25 images shows the channel maps of the residuals after subtracting the best fit model in visibility space and imaging. Contours are drawn at the same level as the data.}
    \label{fig:Channel_map_fits}
\end{figure*}

We compare the synthetic CO model cubes produced from DISC2RADMC to the continuum subtracted visibilities of the CO emission. In the fit we include all channels with CO detected at a 3$\sigma$ level as well as 3 channels on either side of the CO line that do not have a significant detection in order to get the best constraint on the gas inner radius. We omit the channels that have diffuse emission detected at a 3$\sigma$ level at the position of HD 166191 from our fitting procedure, as it could affect the derived disk parameters. The only channel that has this diffuse emission at the position of HD 166191 is the channel at 7.81 km/s as shown in Figure \ref{fig:Channel_map_fits}. We include $R_c$, $\gamma$, and $R_{in}$ as free parameters as well as the disk inclination and position angle and total CO mass. We also fit for the system's LSRK velocity by shifting the CO model cubes by a velocity that is left as a free parameter. The scaling of the CO temperature, with a fixed profile as a function of radial distance from the star is also left as a free parameter. To report the temperature results from the modeling, we report it as $T_0$, which represents the gas temperature at 1 au. We also fit for an offset in R.A. and Dec (dRA and dDec) and leave these values as free parameters. We use GALARIO \citep{galario18} to transform the synthetic CO images at each frequency into visibilities sampled at the same UV points as the observations. The model visibilities are then compared to the real and imaginary visibilities of the data at each frequency channel. HD 166191 was observed in 3 separate observing blocks, giving three sets of visibilities. For each model, we calculate the chi-squared value between each three sets of visibilities separately and then add the three chi-squared values together. This chi-squared value is then used in a Gaussian likelihood function and the emcee package is used to infer the posterior distribution of each of the free parameters with an MCMC \citep{emcee13}. We run the MCMC for 2000 steps with 40 walkers. Before comparing the model visibilities to the data, we re-weight the data visibilities using the CASA task statwt. To ensure that the absolute uncertainties are appropriate, we scale the weights so that the reduced chi-squared value of the best fit model is 1 \citep{Marino21,Hales22}.

 The channel maps showing the CO data and the residuals after subtracting the best fit model are shown in Figure \ref{fig:Channel_map_fits} . The residuals channel maps are computed by subtracting the best fit model visibilities from the data and then re-imaging the cube with tclean using the same imaging parameters described above. The model appears to be a decent fit to the data as most of the channel maps after subtracting the model appear to have no significant residuals, although two have residuals at the 3-6$\sigma$ level. A table of the best fit parameters for the CO model as well as the priors used in the fit are shown in Table \ref{tab:CO_params}. For each of the parameters, we used flat priors. The posterior distributions for each of the model parameters in the CO model are shown in the Appendix. As shown in Figure \ref{fig:Surface_dens}, the CO also appears to be constrained within a stellocentric distance of 20 au, similar to the dust. With the velocity information from the channel maps, the inner radius of the CO is constrained to be 2.21$^{+0.36}_{-0.37}$ au.

 We also compute our best fit CO model for the J=2-1 transition and compare the line flux to the upper limit from the SMA data. We find that the CO J=2-1 integrated line flux from the best fit model is 0.5 Jy km/s. The 3$\sigma$ upper limit from the SMA data on the CO J=2-1 line flux is 1 Jy km/s, which is larger than the line flux from our best fit model. Our best fit model to the CO J=3-2 line from ALMA is therefore consistent with the non-detection of the CO J=2-1 line from the SMA.

 \begin{table}[!htbp]
     \centering
     \caption{CO model parameters}
     \begin{tabular}{c c c}
     
        \hline
        
          Model Parameter& Best fit value & Prior Range  \\
          \hline
          log(M$_{\mathrm{CO}}$) (M$_{\oplus}$)&0.83$^{+0.38}_{-0.24}$  & (-5, 1.5)\\
          $R_c$ (au)& 7.1$^{+1.7}_{-1.5}$ & (1, 60)\\
          $\gamma$& -0.43$^{+0.43}_{-0.64}$& (-2, 2)\\
          $R_{in}$ (au) &2.21$^{+0.36}_{-0.37}$& (0.1, 10)\\
          P.A. ($^{\circ}$) & 167.6$^{+0.2}_{-0.2}$& (100, 200)\\
          inc. ($^{\circ}$) &87.9$^{+0.5}_{-0.4}$ &(10, 90)\\
          V$_{\mathrm{LSRK}}$& 5.41$^{+0.02}_{-0.02}$& (4,9)\\
          $T_0$ (K)& 363$^{+6}_{-9}$& (200, 1000)\\
          \hline

     \end{tabular}
     
           \begin{minipage}{6 cm}
    \vspace{0.2cm}
    
     \textbf{Note.} The prior ranges here show the low end of the prior range and the high end of the prior range used in the MCMC. $T_0$ represents the CO temperature at 1 au.
    \end{minipage}
     \label{tab:CO_params}
 \end{table}

\section{Discussion}

\subsection{What the SiO non-detection implies about the collision}
While we are able to place limits on the total SiO mass in the HD 166191 system using our ALMA observations and assuming the emission is optically thin, this limit likely does not represent the amount of SiO that could be present at the time of our observations as photodissociation could destroy the SiO on short timescales after the collision \citep{Johnson12}. To better constrain the amount of SiO that could have survived since the collision, we first estimate the photodissociation rate of unshielded SiO at a distance of 0.6 au, where the collision is thought to have occurred \citep{Su22}. We use the equation 
\begin{equation}
    k=\int_{0}^{\lambda_0}\sigma(\lambda)I(\lambda)d\lambda
\end{equation}
where $\lambda_0$ is the largest wavelength photon that can dissociate SiO (1500 \r{A} \citep{Tardafar90}), $\sigma(\lambda)$ is the photodissociation cross section of SiO as a function of wavelength and is taken from \cite{Heays17}, and $I(\lambda)$ is the photon-based radiation intensity from HD 166191 at 0.6 au. For $I(\lambda)$, we use the stellar photosphere model shown in Figure \ref{fig:SED_fig}. This calculation gives a photodissociation time of SiO at a distance of 0.6 au from HD 166191 of 0.05 years, which would suggest that any SiO produced from the collision between 2018 and 2020 would have been photodissociated before our observations with ALMA if the SiO was not shielded.

Following the procedure in \cite{Johnson12}, we estimate the amount of SiO necessary to self shield such that the photodissociation time is equal to 6 years, the time between the start of the infrared brightening seen with Spitzer and our ALMA observations. In this estimation, we assume that the SiO is in a ring with central distance $R$ from the star and the half width of the ring is $r$ (see \cite{Johnson12} for a diagram of this scenario). The photodissociation time for SiO molecules in this ring is then increased due to self shielding from UV photons to be 
\begin{equation}
    t_{dis}=\frac{N_{SiO}}{\Phi A_{ring}}
\end{equation}
where $N_{SiO}$ is the number of SiO molecules in the ring, $\Phi$ is the flux of photons that are able to photodissociate SiO and is calculated by removing the $\sigma(\lambda)$ term from the integral in equation 5 and then computing the integral, and $A_{ring}$ is the cross sectional area of the ring and is given by $A_{ring}=2\pi R\times2r$. Here we set $R=0.6$ au and we assume that $r=0.3$ au. We then solve for $N_{SiO}$ such that $t_{dis}$ is equal to 6 years and we find that a total SiO mass of $\sim1\times10^{-4}$ M$_{\oplus}$ would be required to increase the photodissociation time of SiO such that could have survived from the time since the collision to the time of our observations with ALMA. 

This SiO mass is greater than the upper limit calculated from the noise level of our observations and is a more realistic limit on the amount of SiO that could be present if it were produced from the collision seen with Spitzer \citep{Su22}. For comparison, some simulations of the impact that formed the Earth-Moon system estimate that the mass of vapor ejected beyond the Roche radius of Earth was 10-30$\%$ of a lunar mass \citep{Canup04}. Using this value of 10$\%$ of a lunar mass, this would give a vapor mass of 1$\times10^{-3}$ M$_{\oplus}$, which is larger than the upper limit placed by the observations and the photodissociation timescale, potentially suggesting that the collisions observed around HD 166191 did not involve as massive as bodies or as high of relative velocities as those used in that simulation of the Moon forming collision.

\subsection{What is the evolutionary state of the system?}
\begin{figure*}[!htb]
    \centering
    \includegraphics[scale=0.5]{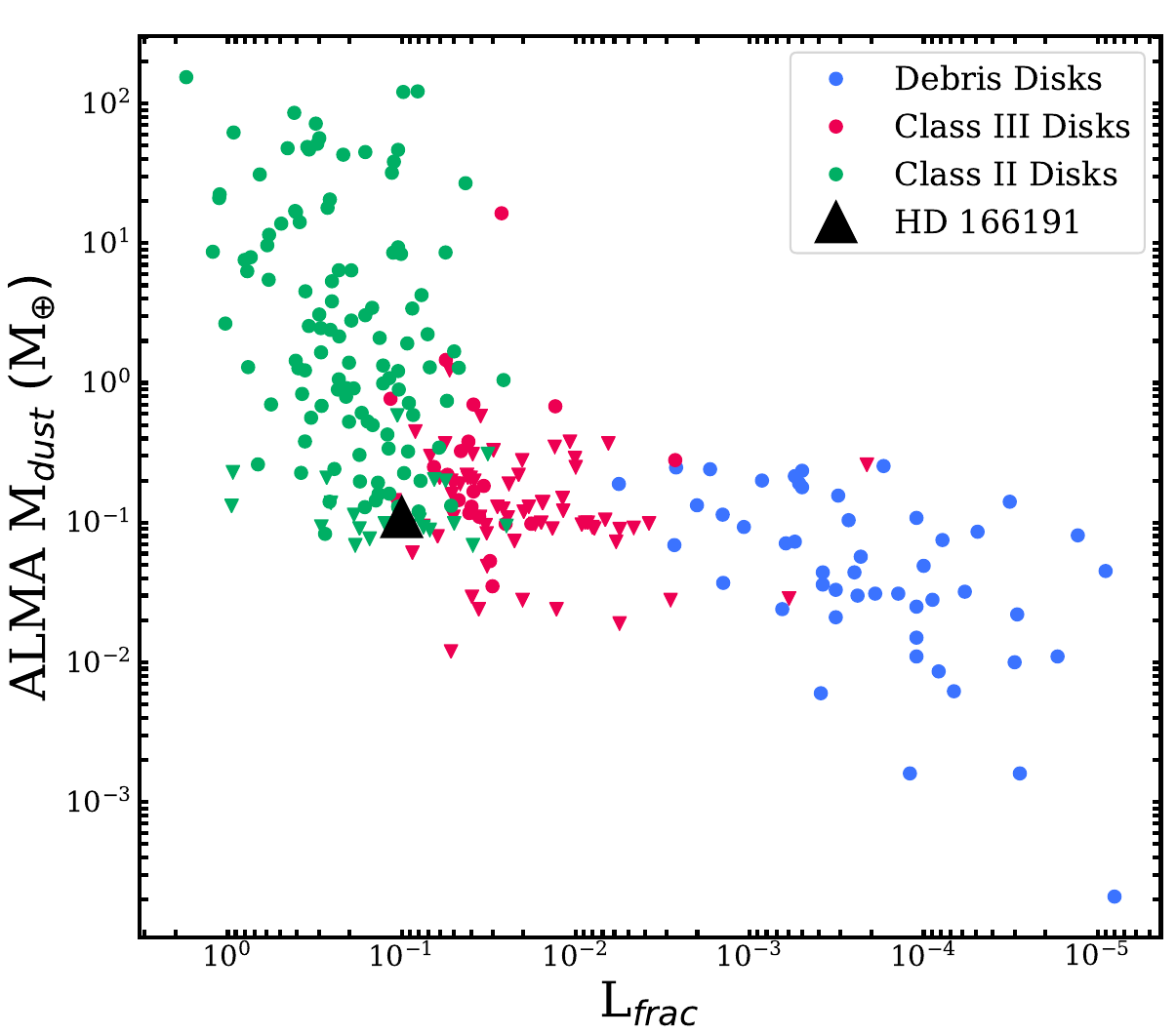}
    \caption{ALMA millimeter dust mass as a function of fractional infrared luminosity for Class II and III disks as well as debris disks from \cite{Michel21}. HD 166191 is shown as a black arrow, since the mass shown here is from the optically thin assumption, which we consider to be a lower limit since radiative transfer models show that the dust mass could be larger. The downward triangles represent systems upper limits and the circles represent systems with ALMA detections.}
    \label{fig:mass_evolution}
\end{figure*}

Given that this system is thought to have a collision in its terrestrial planet zone \citep{Su22}, determining the evolutionary state of this system would help in constraining when and in which environments these types of collisions occur. To try to understand the evolutionary stage of this system, we can use our modeling results of the dust and gas in context with other disks. We can compare dust properties for HD 166191 to other disks in Class II, III, and debris disk stages. Figure \ref{fig:mass_evolution} shows the millimeter dust mass measured for ALMA as a function of fractional infrared luminosity ($L_{IR}$/$L_{*}$) for the sample of Class II and III disks as well as debris disks presented in \cite{Michel21}. HD 166191 is also shown on this Figure, but its dust mass is treated as a lower limit because we find that a marginally optically thick continuum disk can also reproduce the observations, which would give a higher dust mass for HD 166191 than the optically thin assumption. This shows that HD 166191 has a millimeter dust mass lower limit comparable to that of debris disks as well as the low-mass Class II and II disks. In the high dust mass model of HD 166191, however, the dust mass would be larger than any of the debris disks shown in this figure. The SED model of \cite{Kennedy14} also suggests a high dust mass of $\sim$2 M$_{\oplus}$, which is comparable in order of magnitude to our high dust mass model which has a dust mass of 3.6 M$_{\oplus}$. This dust mass derived from our marginally optically thick model here should not be taken as an absolute correct value as this model represents only one instance of an optically thick simulation, given that the computational cost of running the optically thick simulation makes fitting a range of masses of optically thick models difficult. Given the low spectral index from the ALMA and SMA observations ($\sim$2), it is possible that the emission is marginally optically thick and the dust mass is larger than in the optically thin limit, however, we cannot make definitive conclusions about the total dust mass of the system with the current data. The fractional infrared luminosity of HD 166191 of 0.1 \citep{Schneider13} is larger than that of any of the debris disks in the sample but is comparable to Class II disks \citep{Michel21}. However, this may not indicate that this disk is more similar to Class II disks as the high infrared excesses could be caused by large amounts of dust released through collisions, like the one seen with \textit{Spitzer} \citep{Su22}.

If HD 166191 is more similar to protoplanetary or transitional disks than debris disks, it may not be surprising that the millimeter emission would be optically thick. Other compact protoplanetary disks with disk outer radii less than 30 au have also been suggested to possess marginally optically thick continuum emission \citep{Long22,G-A25}. A survey of protoplanetary disks in Taurus similarly found evidence from microwave continuum spectra that the sub-millimeter emission from those disks may be optically thick \citep{Painter25}. Further, an analysis of a sample of 12 protoplanetary disks found that each of the edge on disks had optically thick continuum emission \citep{Villenave20}. Given that the inclination of HD 166191 is found to be edge on from the CO modeling and the total extent of the disk is less than 30 au (see Figure \ref{fig:Surface_dens}), it is possible that HD 166191 also has marginally optically thick continuum emission. Further, \cite{Kennedy14} also suggest that a relatively large optical depth of the disk may be required to produce the fractional infrared excess of the disk around HD 166191.

The best fit CO mass from our model of 6.8$^{+9.4}_{-2.9}$ $M_{\oplus}$ is larger than that of any known gas-rich debris disk (e.g. \citealt{Rebollido22} and references therein), potentially hinting at this disk not being in family with the gas-rich debris disks and more akin to protoplanetary or transition disks. We urge caution in over-interpreting the absolute value of the CO mass from the model here since we only have the one $^{12}$CO line and the line flux from the model appears to rely on some of the assumptions we have made that can currently not be tested (i.e. vertical structure, excitation temperature, surface density profile). Further observations of other CO isotopologues and other CO transitions, potentially at higher angular resolution, would enable a better constraint on the total CO mass in order to make more firm conclusions. The mass currently determined from the model would suggest that this system has more gas than the known gas-rich debris disks, however, with only this one CO line, it is challenging to rule out a secondary scenario for the gas especially given that there appears to have been a large collision in the system. Making the canonical assumption of H$_2$/CO=10,000, and using the best fit CO mass from our model, we get a total gas disk mass for HD 166191 of 0.014$^{+0.021}_{-0.006}$ M$_{\odot}$. For comparison, HD 163296, the Herbig star thought to be co-moving with HD 166191, has a total disk mass of 0.12 M$_{\odot}$ \citep{Pezzotta25}. 

\cite{Kennedy14} also put forward the idea that the outer regions of this disk are gaseous with settled dust while the inner region has begun its transition to a debris disk phase, where the dust is secondary. They propose that the disk may consist of two components, the inner warm component that may be due to collisions from ongoing terrestrial planet formation, and then settled outer component that is gas rich and contains material from the protoplanetary disk. Our observations are generally consistent with this scenario considering the constraint on CO inner radius and estimated CO mass, but we currently cannot be certain that this is the case for this system given the unconfirmed assumptions made in the CO model and the uncertainty in the total dust mass. Nonetheless, HD 166191 appears to be a unique object that is useful for understanding this transition between protoplanetary and debris disks. 

%Our observations also show the compact structure of this disk, with the CO and dust being constrained to within 20 au from the central star. Radial drift has been invoked to explain compact millimeter dust disks \citep{Andrews16,Trapman19,Trapman20}. The definitive signature of radial drift being the cause of compact millimeter dust disks is proposed to be that the CO disk is 4 times more extended than the dust disk \citep{Trapman19}. The similarity in the outer radius of the gas and the dust as shown in Figure \ref{fig:Surface_dens} may suggest then that the compact nature of the millimeter dust is not due to radial drift. The HD 166191 disk may be compact then not because of radial drift but because the disk was born compact \citep{Long22}. 

\subsection{How the collision fits into this system}
If this system is truly a transitional disk with large masses of primordial gas, this would suggest that collisions in the terrestrial planet zone would be occurring early in this system's evolution (an age of $\sim$4 Myr \citep{Hunt23,Hunt24}), while there is still large levels of gas in the outer regions of the disk. \cite{Kennedy14} suggest through SED modeling and the lack of accretion signatures for the system that it has an inner hole, with an inner radius of about 1 au. Our CO modeling places a similar constraint of the inner CO radius being 2.21$^{+0.36}_{-0.37}$ au. The collision found with Spitzer was thought to occur at 0.6 au, which could put this event within this inner radius of the disk \citep{Su22}. That is, inside a region which is gas depleted and thus does not possess sufficient gas densities to dampen the eccentricities and relative velocities of planetesimals, which would prevent destructive collisions from occurring in gas rich regions \citep{Guilera14}. If HD 166191 is host to a transition disk with some remnant primordial gas, then this system would be of interest for understanding how protoplanetary disks transition into debris disks, where the dust is of second generation \citep{Wyatt07, Hughes18}. Regardless of the evolutionary state of the system, our observations combined with the results of \cite{Su22} show that collisions in the terrestrial planet zone of this system are occurring in a stage of evolution where there is circumstellar gas present. 

Given the constraints on the disk geometry we have from the CO fitting, we can compare them with what is known about the system and collision geometry. The geometry of the gas determined from the model fitting suggests that the disk is edge-on with an inclination of $\sim88^{\circ}$ (see Table \ref{tab:CO_params}). One of the collisional indicators seen by \cite{Su22} was two transits of a dust clump, which suggests that orbital plane of the planetesimals that collided were edge on. Our results suggest that the gas orbits roughly in a similar edge on plane as the dust clumps that created the transit signals at optical wavelengths. The vsin($i$) of HD 166191 of 27 km/s, compared to other F-type stars, also suggests that the stellar rotation axis is perpendicular to the line of sight \citep{Potranov18,Su22}, which is generally consistent with the edge on CO disk observed here. The inclination derived from our radiative transfer modeling (87.9$^{+0.5}_{-0.4}$ degrees) differs from that derived from the Gaussian fit to the image (69$^{+9}_{-11}$ degrees) at the 2-3$\sigma$ level. The geometric calculation of the inclination, however, does not account for any vertical structure of the disk, potentially leading to an incorrect inclination value. Further, the radiative transfer modeling of the CO uses the kinematic profile of the CO along with the spatial profile to derive the inclination, unlike the geometric calculation of the inclination. Ultimately, higher angular resolution observations would help better constrain the inclination. 

Another system thought to have undergone a large collision is HD 172555 \citep{Lisse09, Schneidermann21}. HD 172555 also has CO detected with ALMA, however, it is spatially unresolved and thought to be optically thin with a lower CO mass of 1.4$\times10^{-5}$ M$_{\oplus}$ \citep{Schneider13}, in contrast to HD 166191 which is better explained by a higher mass optically thick CO model. Further, the CO in the case of HD 166191 does not appear to be co-spatial with where the collision is thought to have occurred in the system. The high CO mass we get from the model seems to suggest that this system is more gas rich than typical debris disks that possess gas (e.g. \citealt{Moor17,Moor19,Schneidermann21, Rebollido22}), although due to the assumptions in the CO model that affect the mass determination, we cannot firmly rule out the secondary gas scenario for HD 166191. Confirming the high CO mass with CO isotopologues or detecting H$_2$ would be possible future indicators that could support the primordial gas scenario.

\section{Conclusions}
We present SMA and ALMA band 7 observations of the HD 166191 system and our main findings are:
\begin{itemize}
    \item We detect millimeter continuum and CO emission from the HD 166191 disk, while putting an upper limit on SiO.
    \item Unlike the infrared emission from the source, we do not find evidence for variability at millimeter wavelengths between 2014 and 2024.
    \item Modeling of the dust and CO reveals that it is relatively compact and confined to within 20 au from the central star.
    \item The modeling, with some caveats, favors a CO mass that is larger than gas-rich debris disks, potentially being consistent with HD 166191 hosting a evolved protoplanetary or transition disk.
    \item The dust continuum can be reproduced by both an optically thin and moderately optically thick model, making the dust mass uncertain. 
    \item If this system does host large amounts of primordial gas, as hinted at by our results, this would suggest that collisions in the terrestrial planet zone are occurring while this system is still transitioning out of its protoplanetary disk phase of evolution. 
\end{itemize}

KW and CC acknowledge the support from the NASA FINESST program. This work is supported by the National Aeronautics
and Space Administration under Grant No. 80NSSC22K1752
issued through the Mission Directorate. The SMA is a collaborative project of the Smithsonian Astrophysical Observatory (SAO) and the Academica Sinica Institute of Astronomy and Astrophysics (ASIAA). This paper makes use of the following ALMA data: ADS/JAO.ALMA$\#$2023.1.00691.S ALMA is a partnership of ESO (representing its member states), NSF (USA) and NINS (Japan), together with NRC (Canada), NSTC and ASIAA (Taiwan), and KASI (Republic of Korea), in cooperation with the Republic of Chile. The Joint ALMA Observatory is operated by ESO, AUI/NRAO and NAOJ. The National Radio Astronomy Observatory is a facility of the National Science Foundation operated under cooperative agreement by Associated Universities, Inc. We wish to recognize and acknowledge the very significant cultural role and reverence that the summit of Mauna Kea has always had within the indigenous Hawaiian community.  We are most fortunate to have the opportunity to conduct observations from this mountain. We also acknowledge helpful comments on the initial proposal and data from Cicero Lu.

\software{
This research has made use of the following software projects:
    \href{https://astropy.org/}{Astropy} \citep{Astropy13,astropy18,Astropy22},
    \href{https://matplotlib.org/}{Matplotlib} \citep{matplotlib07},
    \href{http://www.numpy.org/}{NumPy} and \href{https://scipy.org/}{SciPy} \citep{numpy07}
    and
    the NASA's Astrophysics Data System.
}
\appendix

\begin{figure*}[!htpb]
    \centering
    \includegraphics[scale=0.71]{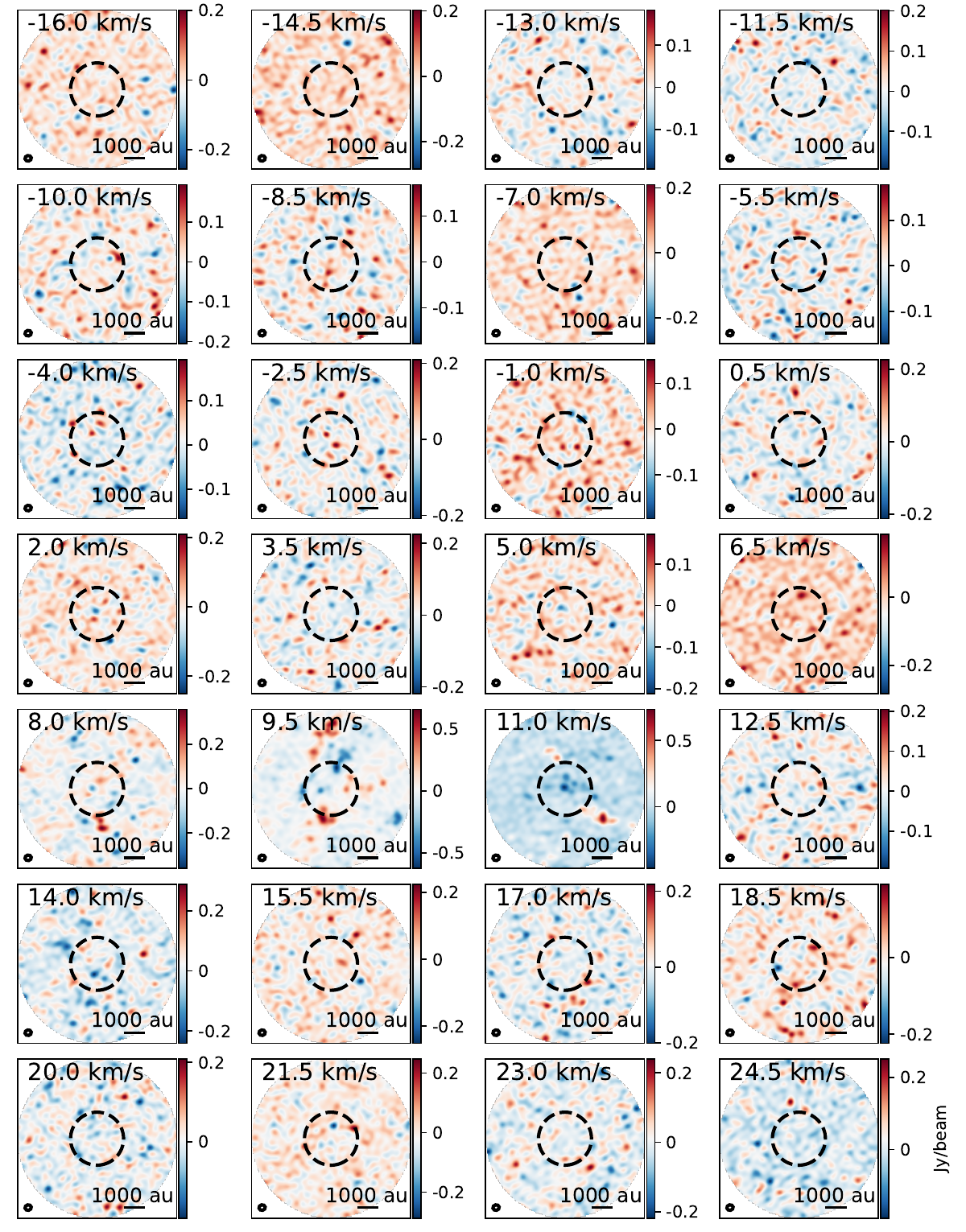}
    \caption{Channel maps of HD 166191 from the SMA covering the CO J=2-1 line. No detection of CO from the disk is found in the channel maps, however, emission from the surrounding cloud is detected. The black ellipse in the channel maps shows the synthesized beam. The black dashed circle in the center of the channel maps shows the ALMA field of view displayed in Figure \ref{fig:diffuse_CO}. The velocities shown in the channel maps are LSRK velocities. }
    \label{fig:SMA_CO}
\end{figure*}

\begin{figure*}
    \centering
    \includegraphics[scale=0.4]{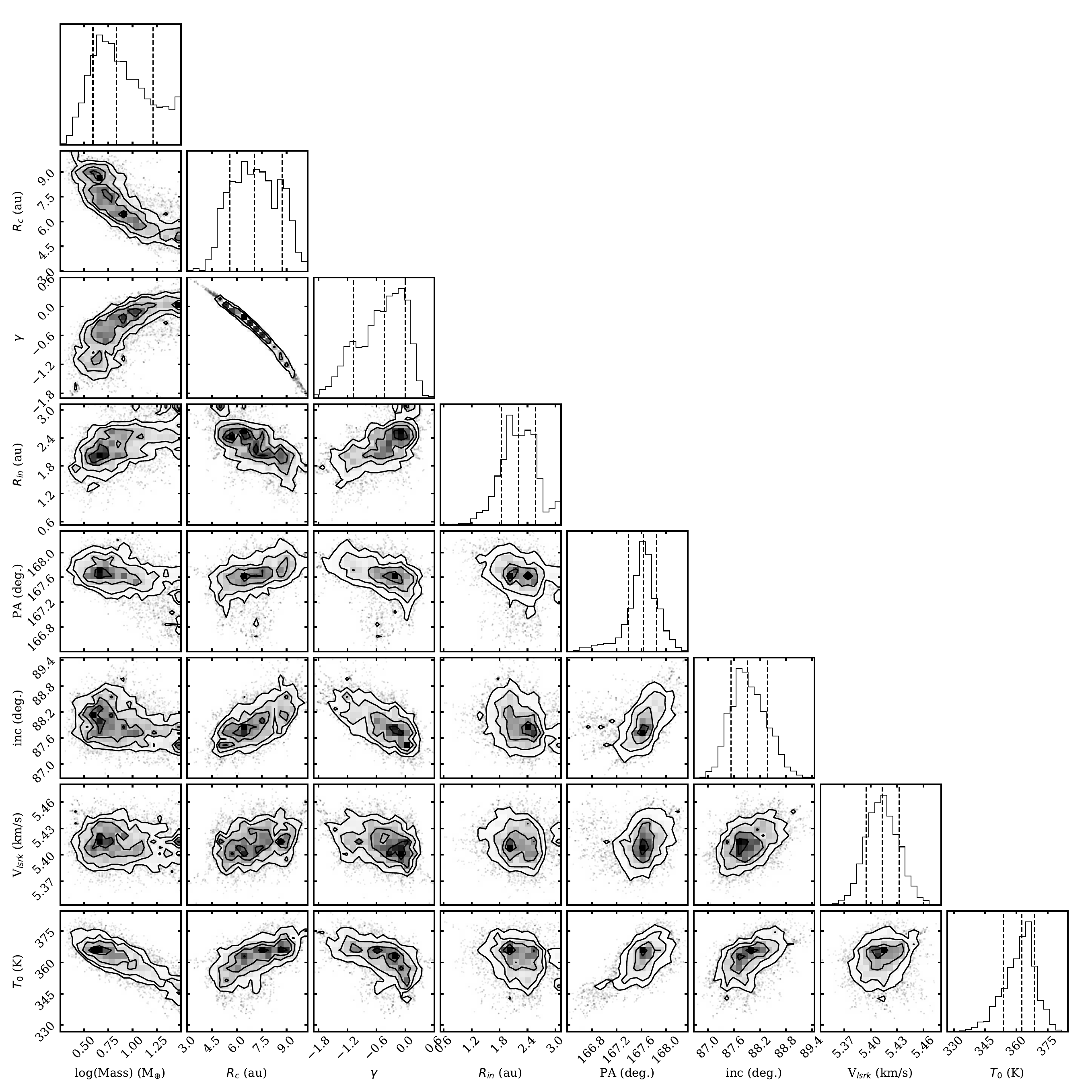}
    \caption{Posterior distributions of each of the parameters from the radiative transfer fitting to the CO visibility data. The dashed lines represent the 16th and 84th percentiles of the posterior distribution as well as the median.}
    \label{fig:Corner_CO}
\end{figure*}

\begin{figure}
    \centering
    \includegraphics[scale=0.6]{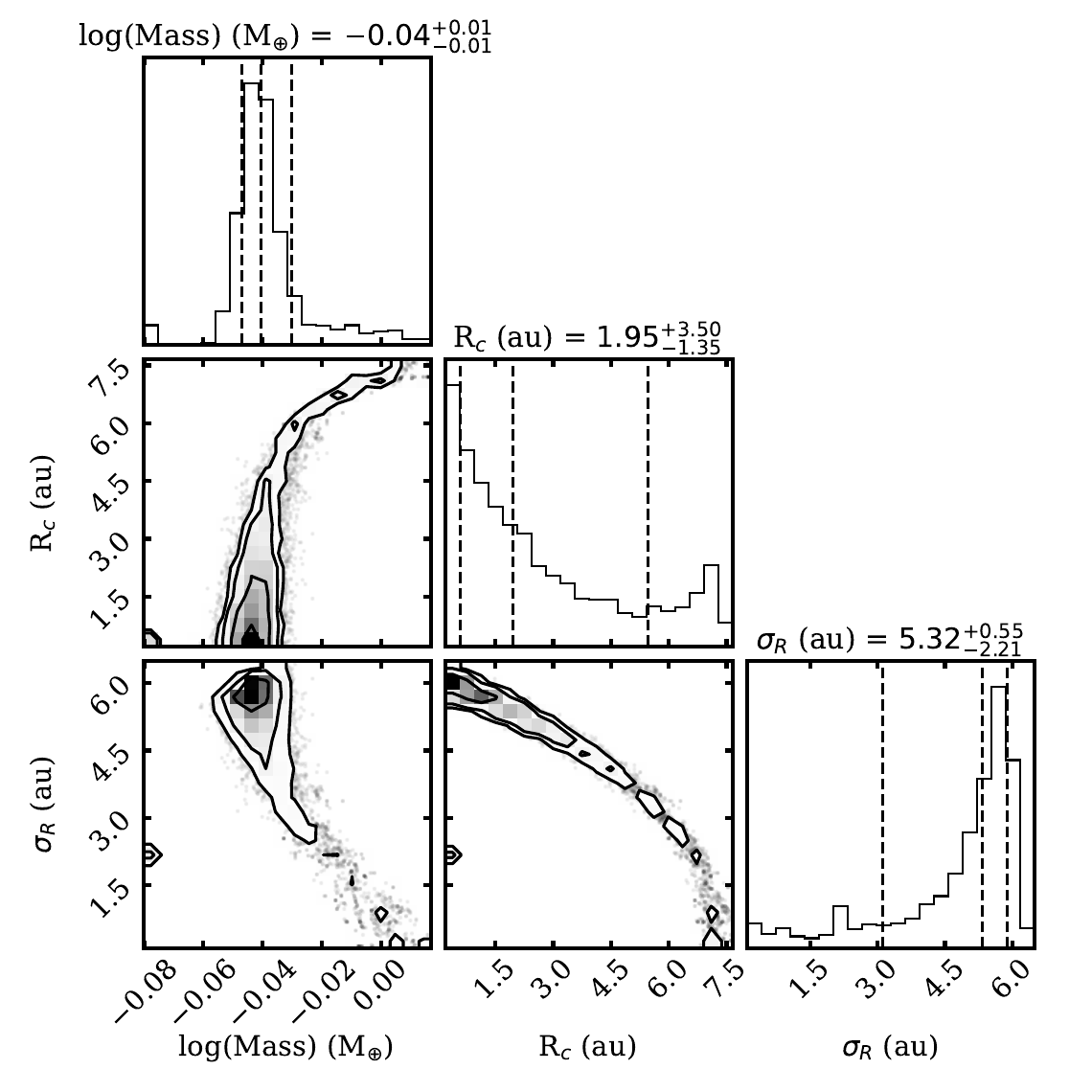}
    \caption{Posterior distributions for the low mass dust model. The dashed lines represent the same quantities as in Figure \ref{fig:Corner_CO}.}
    \label{fig:dust_low_corner}
\end{figure}

\begin{figure}
    \centering
    \includegraphics[scale=0.6]{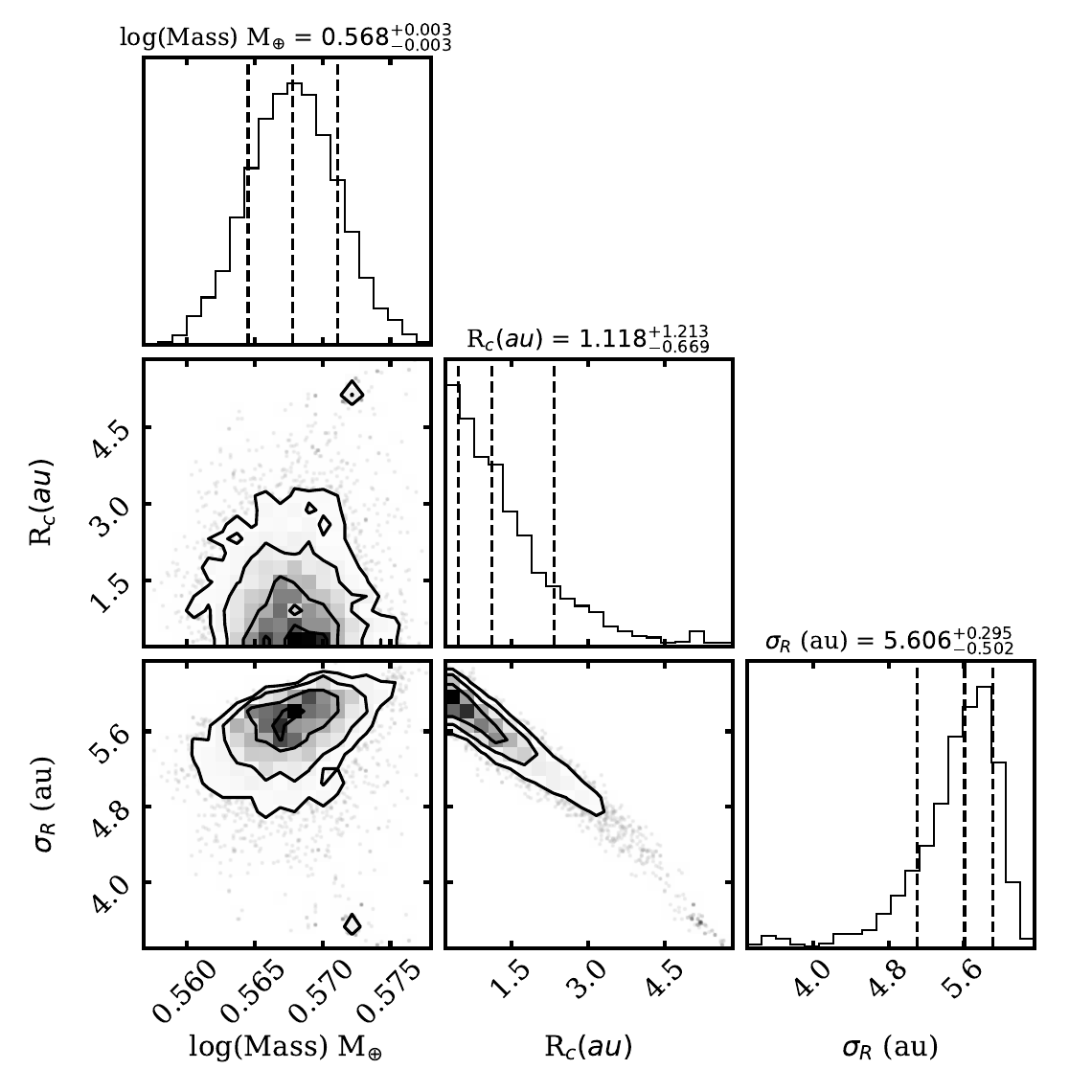}
    \caption{Same as Figure \ref{fig:dust_low_corner} but for the high dust mass model that is marginally optically thick.}
    \label{fig:dust_high_corner}
\end{figure}

%%%%%%%%%%%%%%%%%%%%%%%%%%%%%%%%%%%%%%%%%%%%%%%%
%				Bibliography
%%%%%%%%%%%%%%%%%%%%%%%%%%%%%%%%%%%%%%%%%%%%%%%%

\bibliographystyle{aasjournal}
\bibliography{mybib}

%%%%%%%%%%%%%%%%%%%%%%%%%%%%%%%%%%%%%%%%%%%%%%%%
%				  Figures
%%%%%%%%%%%%%%%%%%%%%%%%%%%%%%%%%%%%%%%%%%%%%%%%

%\begin{figure}
%    \centering
%    \includegraphics[width=0.47\textwidth]{figs/bgps4029_12m_spec.pdf}
%\caption{
%{\it Top:} \hcop, \htcop, and \hceop\ $J=1-0$ spectra from the ARO 12m showing the characteristic red-shifted self-absorption profile.
%The large $70\arcsec$ 12m beam creates significant beam dilution, partly leveling out the peak asymmetry.
%The brightness temperature is on the main-beam scale, and note that the \hceop\ data is shown in mK.
%}
%\label{fig:TwelveMeter}
%\end{figure}

%%%%%%%%%%%%%%%%%%%%%%%%%%%%%%%%%%%%%%%%%%%%%%%%
%				  Tables
%%%%%%%%%%%%%%%%%%%%%%%%%%%%%%%%%%%%%%%%%%%%%%%%

%\clearpage

%\input{tabs/htcop_props.tex}  % tab:HtcopProps

\end{document}